\documentclass{egpubl}
\usepackage{eurovis2022}

\JournalSubmission    

\usepackage[T1]{fontenc}
\usepackage{dfadobe}  

\usepackage{cite}  
\BibtexOrBiblatex
\electronicVersion
\PrintedOrElectronic
\ifpdf \usepackage[pdftex]{graphicx} \pdfcompresslevel=9
\else \usepackage[dvips]{graphicx} \fi

\usepackage{egweblnk}

\graphicspath{{figures/}{pictures/}{images/}{./}} 

\usepackage{microtype}                 
\PassOptionsToPackage{warn}{textcomp}  
\usepackage{textcomp}                  
\usepackage{mathptmx}                  
\usepackage{times}                     
\usepackage{booktabs}
\usepackage{multirow}
\usepackage{makecell}
\usepackage{xcolor}
\usepackage{amsmath}
\usepackage{amssymb}
\usepackage{mathtools}
\usepackage{bbm}
\usepackage{bm}
\usepackage{algorithm}
\usepackage{algpseudocode}
\usepackage{array}
\newcolumntype{L}[1]{>{\raggedright\let\newline\\\arraybackslash\hspace{0pt}}m{#1}}
\newcolumntype{C}[1]{>{\centering\let\newline\\\arraybackslash\hspace{0pt}}m{#1}}
\newcolumntype{R}[1]{>{\raggedleft\let\newline\\\arraybackslash\hspace{0pt}}m{#1}}
\usepackage[caption=false,font=footnotesize,labelfont=sf,textfont=sf]{subfig}
\usepackage[percent]{overpic}
\usepackage[export]{adjustbox}
\usepackage{tikz}
\usepackage{units}
\usepackage[shortcuts]{extdash}
\usepackage{pdfpages}

\newcommand{\etAl}{\textit{et~al.}}

\newcommand{\R}{\mathbb{R}}
\newcommand{\N}{\mathbb{N}}

\newcommand{\changed}[1]{#1}

\algnewcommand\algorithmicparameters{\textbf{Parameters:}}
\algnewcommand\Parameters{\item[\algorithmicparameters]}
\algnewcommand\algorithmicinput{\textbf{Input:}}
\algnewcommand\Input{\item[\algorithmicinput]}
\algnewcommand\algorithmicoutput{\textbf{Output:}}
\algnewcommand\Output{\item[\algorithmicoutput]}
\algnewcommand\algorithmicforward{\textbf{Forward:}}
\algnewcommand\Forward{\item[\algorithmicforward]}
\algnewcommand\algorithmicbackward{\textbf{Backward:}}
\algnewcommand\Backward{\item[\algorithmicbackward]}

\title[fV-SRN for DVR]{Fast Neural Representations for Direct Volume Rendering}

\author[S. Weiss \& P. Herm{\"u}ller \& R. Westermann]
{\parbox{\textwidth}{\centering Sebastian Weiss\orcid{0000-0003-4399-3180}
        and Philipp Herm{\"u}ller\orcid{0000-0002-8743-7488} 
        and R{\"u}diger Westermann\orcid{0000-0002-3394-0731}
        }
        \\
{\parbox{\textwidth}{\centering Technical University of Munich, Germany}}
}

\begin{document}


\maketitle

\begin{abstract}
Despite the potential of neural scene representations to effectively compress 3D scalar fields at high reconstruction quality, the computational complexity of the training and data reconstruction step using scene representation networks limits their use in practical applications. In this paper, we analyze whether scene representation networks can be modified to reduce these limitations and whether such architectures can also be used for temporal reconstruction tasks. We propose a novel design of scene representation networks using GPU tensor cores to integrate the reconstruction seamlessly into on-chip raytracing kernels, \changed{and compare the quality and performance of this network to alternative network- and non-network-based compression schemes. The results indicate competitive quality of our design at high compression rates, and significantly faster decoding times and lower memory consumption during data reconstruction.}
\changed{We investigate how density gradients can be computed using the network and show an extension where density, gradient and curvature are predicted jointly.}
As an alternative to spatial super-resolution approaches for time-varying fields, we propose a solution that builds upon latent-space interpolation to enable random access reconstruction at arbitrary granularity. We summarize our findings in the form of an assessment of the strengths and limitations of scene representation networks \changed{for compression domain volume rendering, 
and outline future research directions}. Source code: \url{https://github.com/shamanDevel/fV-SRN}\\

\printccsdesc   
\end{abstract}

\section{Introduction}

Learning-based lossy compression schemes for 3D scalar fields using neural networks have been proposed recently. While first approaches have leveraged the capabilities of such networks to learn general properties of scientific fields and use this knowledge for spatial and temporal super-resolution~\cite{zhou2017volume,han2020ssr,guo2020ssr,han2019tsr}, Lu~\etAl~\cite{lu2021compressive} have focused on the use of Scene Representation Networks (SRNs)~\cite{park2019deepsdf,sitzmann2019scene,mildenhall2020nerf} that overfit to a specific dataset to achieve improved compression rates. 

SRNs were introduced as a compact encoding of (colored) surface models. They 
replace the initial model representation with a learned function that maps from domain locations to surface points. 
SRNs are modeled as fully connected networks where the scene is encoded in the weights of the hidden layers.
This scene encoding---the so-called latent-space representation---can be trained from images of the initial object via a differentiable ray-marcher, or in object-space using sampled points that are classified as inside or outside the surface.
Since SRNs allow for direct access of the encoded model at arbitrary domain points, ray-marching can work on the compact representation without having to decode the initial object. 

\changed{Lu~\etAl~\cite{lu2021compressive} introduced \emph{neurcomp}, an SRN where the mapping function has been trained to yield density samples instead of surface points.} We subsequently refer to an SRN that predicts density samples as Volume Representation Network (V-SRN).
By using a V-SRN, a ray-marcher can sample directly from the compact latent-space representation, and does not require to decode the initial volume beforehand. However, at every sample point along the view-rays, a deep network is called to infer the density sample.


Since SRNs are implemented using generic frameworks like PyTorch or Tensorflow 
where the basic building block is a network layer, intermediate states of each layer need to be written to global memory to make it available to the next layer. Thus, the evaluation becomes heavily memory-bound when deep networks consisting of multiple layers are used. 
Due to this reason, direct volume rendering using V-SRN is currently limited to non-interactive applications, with framerates that are significantly below what can be achieved on the initial data. Furthermore, the size of the networks that are used to generate the model representation drastically increases the training times. 

\begin{figure*}
    \centering
    \includegraphics[width=\linewidth]{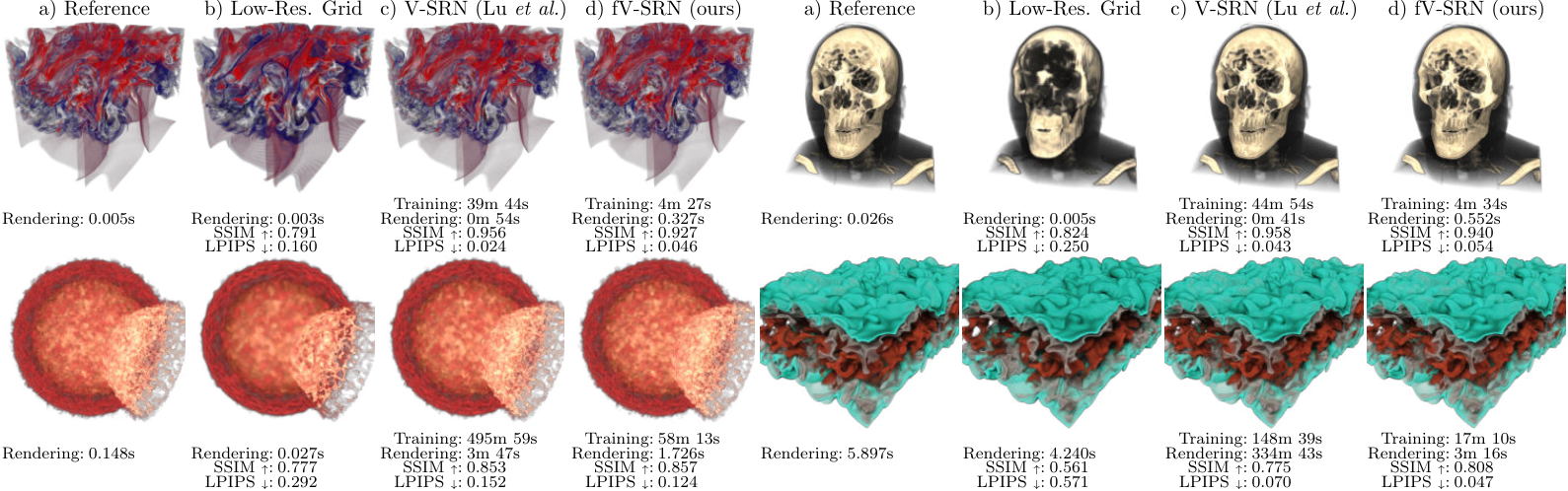}%
    \caption{
    Performance and quality comparison of volume representation networks (SRNs). a) The original datasets. b) Low-pass filtered versions, subsampled to meet the selected compression rate. c) \changed{neurcomp} by Lu~\etAl~\cite{lu2021compressive}. d) Our proposed fV-SRN. \changed{From left to right, top to bottom: Richtmyer-Meshkov (RM), Skull (each of resolution $256^3$), Ejecta ($1024^3$), Jet ($512\times 336\times 768$, courtesy of Lu~\etAl). The same network configuration was used for all datasets. Compression rates for b) to d): $1:32$ for RM and Skull, $1:254$ for Jet, $1:2048$ for Ejecta, including network weights and latent grid.
    RM, Skull, Ejecta rendered using DVR, Jet using Monte-Carlo path tracing with $256$ spp to a screen of resolution $1024^2$. RM and Skull trained using $256^3$ samples, Jet and Ejecta using $512^3$ and $1024^3$ samples, respectively.}
    }
    \label{fig:compressionTeaser}
\end{figure*}

\autoref{fig:compressionTeaser} demonstrates the aforementioned properties for different datasets and a given memory budget of roughly $3\%$ \changed{to $0.4\%$} of the memory that is required by the original dataset.
Given the internal format of the network weights, which is set to 16-bit half-precision floating-point values in the current examples, V-SRN automatically determines the internal network layout so that the memory budget is not exceeded.
Compared to the initial datasets in (a), V-SRN in (c) with 18 layers (8 residual blocks) and 128 channels shows high reconstruction fidelity at the given compression rate. Compared to low-pass filtered versions of the original datasets (b), which are resampled to a resolution that matches the memory budget, even fine structures are well preserved. However, the rendering times are between 41 seconds \changed{to multiple minutes}, and training times range from 39 minutes \changed{to multiple hours}.

\subsection{Contribution}
In this work, we demonstrate that the efficiency of V-SRN for volume rendering can be significantly improved, both with respect to training and data reconstruction. We achieve this by a novel compact network design called fV-SNR, which effectively utilizes the GPU TensorCores and uses a trained volumetric grid of latent features as additional network inputs. This enables fast training and \changed{significantly faster rendering from the compressed representation than prior work.} 


\changed{We compare fV-SRN to \emph{neurcomp} by Lu~\etAl~\cite{lu2021compressive} as well as non-network-based compression schemes TThresh~\cite{ballester2019tthresh} and cudaCompress~\cite{treib2012interactive,treib2012turbulence} regarding compression ratio and reconstruction speed. The results indicate that fV-SRN is significantly faster than \emph{neurcomp} at similar compression rates, achieves similar compression rates than TThresh at significantly lower reconstruction times, and significantly outperforms cudaCompress in terms of compression ratio at similar decoding speed. Furthermore, since fV-SRN can render directly from the compressed representation, no additional memory is required at rendering time. 
Building upon the strengths of fV-SRN, we introduce the extension of fV-SRN to predict scalar field values as well as derived quantities like gradients and curvature estimates.}



We further demonstrate the use of fV-SRN for temporal super-resolution tasks, \changed{to perform smooth, yet structure-preserving interpolation between given volumetric datasets at two consecutive timesteps. This enables to reduce the number of timesteps that need to be stored out of a running simulation. We analyze the possibility of latent-space interpolation to perform this task, and demonstrate that the restriction of available super-resolution schemes like TSR~\cite{han2019tsr} and STNet~\cite{han2021stnet} to obtain interpolations only at a pre-defined discrete set of timesteps can be overcome.}
Our specific contributions are: 
\begin{itemize}
    \item The design and implementation of a fast variant of V-SRN (fV\=/SRN) using a volumetric latent grid and running completely on fast on-chip memory.
    \item \changed{An extension of fV-SRN to jointly predict a scalar quantity as well as the gradient and curvature at the given input position.}
    \item A temporal super-resolution fV\=/SRN using latent-space interpolation as a means for feature-preserving reconstruction of time-sequences at arbitrary temporal resolution.
\end{itemize}


\changed{In an ablation study we shed light on the design decisions and training methodology, and we perform a number of experiments to demonstrate the specific properties of fV\=/SRN. Its quality and performance is compared to state-of-the-art compression schemes targeting direct volume rendering applications.}
Our experiments include qualitative and quantitative evaluations, which indicate high compression quality even when small networks are used. fV\=/SRN can be integrated seamlessly into ray- and path-tracing kernels, and---compared to \changed{\emph{neurcomp}}---improves the rendering performance about two orders of magnitude (between $76\times$ and $165\times$) and the performance of the training process about a factor of $9\times$, see \autoref{fig:compressionTeaser}.
Due to the use of a low-resolution latent grid, temporal super-resolution between given instances in time can be used even for large time-varying sequences (\autoref{sec:unsteady}).



\section{Related Work}
There is a vast body of literature on compression schemes for volumetric fields and scene representation networks, and a comprehensive review is beyond the scope of this paper. However, for thorough overviews and discussions of the most recent works in these fields let us refer to the articles by Balsa Rodr\'{\i}guez et al.~\cite{rodriguez2013survey}, Beyer et al.~\cite{10.2312:eurovisstar.20141175}, Hoang et al.~\cite{Hoang2021EfficientAF}, and Tewari et al.~\cite{tewari2020state}.

\paragraph*{Lossy Volume Compression Schemes}
Our approach, since it attempts to further improve the compressive neural volume representation by Lu~\etAl~\cite{lu2021compressive}, falls into the category of lossy compression schemes for volumetric scalar fields. Previous studies in this field have utilized quantization schemes to represent contiguous data blocks by a single index \changed{or a sparse combination} of \changed{learned} representative values~\cite{Schneider2003compressiondomain, fout2007transform, gobbetti2012covra, guthe2016variable}, or lossy curve fittings like the popular SZ compression algorithm~\cite{di2016fast,kai2020sz}. Transform coding-based schemes~\cite{yeo1995volume,westermann1995compression,lee2008fast} make use, in particular, of the discrete cosine and wavelet transforms. They try to transform the data into a basis in which only few coefficients are relevant while many others can be removed. 
More recently, Ballester-Ripoll~\etAl~\cite{ballester2019tthresh} introduce tensor decomposition to achieve extremely high compression rates \changed{exceeding 1:1000}.

For interactive applications, \changed{methods combining transform coding-based schemes and other techniques listed above} are often applied brick-wise and embedded into streaming pipelines~\cite{treib2012turbulence, reichl2013visualization,diaz2020interactive}. \changed{They achieve significantly smaller compression ratios as, e.g., TThresh, for the sake of efficient GPU decoding.}
\changed{For example, Marton~\etAl~\cite{marton2019compression} present a rendering pipeline capable of decompressing over 10Gvoxels/second while reporting a compression ratio of 1:64 (0.5 bits per sample on floating point data).}
\changed{In our work we target the high compression rates achieved by offline schemes like TThresh, while still being able to render images of large volumes from the compressed representation within a second.}


Fixed-rate texture compression formats such as \changed{ASTC}, S3TC and variants ~\cite{iourcha1999system,fenney2003texture,nystad2012adaptive} are implemented directly by the graphics hardware. This means that rendering, including hardware-supported interpolation, is possible directly from the compressed stream. However, the fixed-rate stage allows little or no control over the quality vs. compression rate trade-off. 

\paragraph*{Deep Learning for Scientific Data Compression}
With the success of convolutional neural networks, deep learning methods have started to see applications in visualization tasks.
Early works use super-resolution networks to upscale the data if either storing the high-resolution data is too expensive (3D spatial data~\cite{zhou2017volume,han2020ssr,guo2020ssr}, temporal data~\cite{han2019tsr}, spatiotemporal data~\cite{han2021stnet}) or the rendering process is too expensive (2D data~\cite{weiss2019isosuperres,weiss2020adaptivesampling}).
Sahoo and Berger~\cite{sahoo2021vectorsuperres} extended 3D super-resolution for vector fields by introducing a loss function that penalizes differences in traced streamlines, instead of only point-wise differences. 
The most recent approach by Lu~\etAl~\cite{lu2021compressive} and Wurster~\etAl~\cite{wurster2021deep} utilize SRNs to learn a compact mapping from domain positions to scalar field values.   

Berger~\etAl~\cite{berger2018generative} and Gavrilescu~\cite{gavrilescu2020supervised} avoid the rendering process completely and train a network that directly predicts the rendered image from camera and transfer function parameters.
This results in a compact representation of the data in the network weights from which the image can be directly predicted, but is limited concerning the generalization to new views or transfer functions if the training data does not provide this specific combination.
Super-resolution methods for 3D spatial data or temporal data, on the other hand, are fixed on a regular grid in space (or time) due to the use of convolutional (recurrent) networks. Therefore, they do not allow for free interpolation and require the decompression of a whole block before rendering.

\paragraph*{Scene Representation Networks}
Scene Representation Networks (SRNs) address the above issues. By directly mapping a spatiotemporal position to the data value, random access is possible, as opposed to grid-based super-resolution methods. This also allows to freely move the camera during testing.

SRNs were first introduced for representing 3D opaque meshes, either as occupancy grids~\cite{mescheder2019occupancy,martel2021acorn} or signed distance field~\cite{takikawa2021neural,davies2020effectiveness,lei2021learning,chabra2020deep}. In these methods, the networks were trained in world-space, that is, from pairs of position to data value.
This principle was also adopted in the VIS area by Lu~\etAl~\cite{lu2021compressive}. The large network used in this work, however, makes interactive rendering infeasible.
For image-space training, that is, training from images through the rendering process, SRNs were first introduced for 3D reconstruction~\cite{sitzmann2019scene}, including \textit{NeRF} by Mildenhall~\etAl~\cite{mildenhall2020nerf,tancik2020fourier}.
Further improvements to these methods include reduction of aliasing artifacts (Mip-NeRF)~\cite{barron2021mip}, \changed{amortized} speed improvements by caching \changed{already evaluated samples} (FastNeRF)~\cite{garbin2021fastnerf}, or incorporation of lighting effects~\cite{srinivasan2021nerv}. 

\changed{Let us remark that in the mentioned scenarios the networks are trained to predict single surfaces from images, i.e., a computer vision task. This allows for significantly larger step sizes, or the use of sparse latent grids where only regions close to the surface are resolved at high resolution~\cite{takikawa2021neural,yu2021plenoctrees}. In the extreme case, a network is completely replaced by a learned sparse voxel representation~\cite{yu2021plenoxels}.
These approaches cannot be transferred to the direct volume rendering scenario addressed in our work, were a surface might not be given or is permanently changed by interactively selecting iso-contours in the volumetric scalar field.}
\changed{Nevertheless, we see potential for future transfer of techniques between both worlds, for example, by integrating the proposed custom TensorCore kernels into computer vision tasks, or extending fV-SRN with aliasing-reducing techniques inspired by Mip-NeRF.}


Regarding dynamic scenes, Park~\etAl~\cite{park2019deepsdf} (modeling SDFs) and Chen and Zhang~\cite{chen2019learning} (modeling occupancy grids) introduce a latent vector that allows interpolating between different models. This is the basis for the time interpolation described in \autoref{sec:unsteady}.
Alternatively, Pumarola~\etAl~\cite{pumarola2021d} introduce a second network that models affine transformations from a base model.

\section{Scene Representation Networks}\label{sec:basic}

Let $V$ be a 3D multi-parameter field, i.e., a mapping $\R^3 \rightarrow \R^D$ that assigns to each point in a given domain a set of $D$ dependent parameters. 
In this work, we focus on 3D scalar fields ($D = 1$) and color fields ($D = 4$), where at each domain point either a scalar density value is given or an RGB$\alpha$ sample has been generated via a transfer function (TF) mapping. 

SRNs~\cite{mildenhall2020nerf,tancik2020fourier} encode and compress the field $V$ via a neural network comprised of fully-connected layers. The network takes a domain position as input and predicts the density or color at that position, i.e., a mapping $V_\Theta : \R^3\rightarrow\R^D$.
In detail, let $\bm{v}_0=\bm{p}\in\R^3$ be the input position. Then, layer $i$ of the network is computed as $\bm{v}_{i+1}=a(W_i \bm{v}_i + \bm{b}_i)$, where $W_i$ is the layer weight matrix, $\bm{b}_i$ the bias vector and $a(\cdot)$ the element-wise activation function. The number of layers is denoted by $l$. The output of the last layer $\bm{v}_l\in\R^D$ is the final network output. The intermediate states $\bm{v}_1,...,\bm{v}_{l-1}$ are of size $\R^c$ with $c$ being the number of hidden channels of the network. The matrices $W_i, \bm{b}_i$ are the trainable parameters of the network.

Since the network processes each input position independently of the other inputs, the volumetric field can be decoded at arbitrary positions only where needed, i.e. along the ray during direct volume raycasting. In practice, batches of thousands of positions are processed in parallel.
In the spirit of previous SRNs~\cite{chen2019learning,mescheder2019occupancy,park2019deepsdf}, \changed{\emph{neurcomp}} by Lu~\etAl~\cite{lu2021compressive} is trained in world-space, using training pairs of position and density $(\bm{x},\bm{v})$. Let us refer to \autoref{sec:steady} for a study of the most relevant network parameters and the details of the training method.


As shown by Mildenhall~\etAl~\cite{tancik2020fourier} and Tancik~\etAl~\cite{tancik2020fourier}, when the SRN is trained only with positional input $\bm{p}=(x,y,z)$ and corresponding density or color output, the mapping function cannot faithfully represent high-frequency features in the data.
To avoid this shortcoming, so-called Fourier features are used to lift 3D positions to a higher-dimensional space before sending the input to the network. In this way, the spread between spatially close positions is increased, and positional variations of the output values are emphasized.

Let $\bm{p}\in\R^3$ and $m\in\N$, respectively, be the input positions to the network and the desired number of Fourier features that should be used (see \autoref{sec:fastSRN:TC} for a discussion of how to chose $m$).
Then, a matrix $F\in\R^{m,3}$ -- the so-called Fourier matrix -- is defined. 
Mildenhall~\etAl~\cite{mildenhall2020nerf} propose to construct the Fourier matrix based on diagonal matrices of powers of two $\mathbbm{1}^3$, i.e., 
\begin{equation}
    F_\text{NeRF} = 2\pi\left[ 2^0\cdot\mathbbm{1}^3, 2^1\cdot\mathbbm{1}^3, \cdots 2^{L-1}\cdot\mathbbm{1}^3 \right] \in \R^{3L,3} ,
\end{equation}
where $m=3L$.
The matrix is fixed before the training process and not part of the trainable parameters.
The inputs to the network are then enriched via vector concatenation as  
\begin{equation}
    \bm{v}_\text{fourier} = \bm{v} \oplus \sin(F\bm{v}) \oplus \cos(F\bm{v}),
\end{equation}
where $\oplus$ indicates the concatenation operation.
%

Alternatively, Tancik~\etAl~\cite{tancik2020fourier} reported better reconstruction quality when using random Fourier features, where the entries of the matrix $F$ are sampled from $\mathcal{N}(0;(2\pi\sigma)^2)$ using the dataset-dependent hyperparameter $\sigma$. In our experiments (\autoref{sec:steady:fourierimportance}), however, we could not observe these improvements and, therefore, follow the construction proposed by Mildenhall~\etAl~\cite{mildenhall2020nerf}.
In contrast to \changed{\emph{neurcomp}}, which does not make use of Fourier features, we observed a significant enhancement of the networks' learning skills when incorporating these features.

\section{Fast Volumetric SRN (fV-SRN)}\label{sec:fastSRN}

When using SRNs, the main computational bottleneck is the evaluation of the network to infer a data sample at a given domain location. In frameworks like PyTorch or Tensorflow, the basic building block a network is composed of is a single linear layer. On recent GPUs, when a layer is evaluated, the inputs and weights are loaded from global memory, updated, and the results are written back to global memory to make them available to the next layer. In direct volume rendering, if 100 steps along a ray are taken and the SRN consists of 7 layers, this amounts to 700 layer invocations and global memory read and write operations.
\changed{In the following, we show that it is possible to completely avoid loading and storing the intermediate results to global memory by fusing the network into a single CUDA kernel and following certain size constraints as detailed below. This idea was previously applied for radiance caching in Monte-Carlo path tracing~\cite{mueller2021realtime}, but is extended here by the latent grid (see \autoref{sec:fastSRN:latentgrid}) and by avoiding all global memory access within the network layers (see \autoref{sec:fastSRN:TC}).} This gives rise to a speedup of up to $16.8\times$ of our custom CUDA TensorCore implementation, compared to a native PyTorch~\cite{PyTorch2019} implementation, for the same V-SRN network architecture, see \autoref{sec:steady:performance}.

\subsection{Custom Inference via TensorCores}\label{sec:fastSRN:TC}
NVIDIA GPUs expose 64kB of fast on-chip memory per multiprocessor that is magnitudes faster than global memory (GBs of memory shared across all multiprocessors).
These 64kB are divided into 48kB freely accessible shared memory and 16kB of L1-cache.
Furthermore, the tensor core (TC) units on modern GPUs provide warp-synchronous operations to speed up matrix-matrix multiplications by a factor of $6\times$~\cite{markidis2018nvidia}. A warp is a group of 32 threads that are executed in lock-step on a single multiprocessor. 
The core operation of the TC units is -- for our purpose -- a matrix-matrix multiplication of $16 \times 16$ matrices of 16-bit half-precision floats $D=AB+C$. 
Each thread holds a part of the input and the output matrices in registers and computes a part of the matrix multiplication. The TC API comes with three main limitations: (a) matrix sizes must be a multiple of $16$, (b) inputs and outputs can only be loaded from and stored to shared \changed{or global memory, not registers}, (c) all 32 threads of the warp must execute the same code.

When evaluating an SRN, each layer computes $y=Wx+b$, where $W$ is the weight matrix, $x$ the input state vector, $b$ the bias vector, and $y$ the output state. To use the TC units as described above and regarding constraint (a), however, $x$ must be a matrix with 16 columns. The first idea is to batch the evaluation so that the 32 threads per warp calculate 16 rays. Then, however, half of the threads are idle in operations like TF evaluation. Therefore, we map one thread to one ray, and block the matrix multiplication $Y=WX+B$ in the following way (exemplary for 48 channels per layer):
The matrix $W\in\R^{48,48}$ is split into $3\times 3$ blocks, and $X,Y\in\R^{48,32}$ are split into $3\times 2$ blocks, each block of shape $16\times 16$. The bias $B\in\R^{48,32}$ is broadcasted from the bias vector $b\in\R^{48}$ by setting the column stride to zero.
In total, a layer evaluation of $48$ channels with $32$ rays requires $6$ invocations to the TC units.

Constraint (b) indicates that the weights and biases of the hidden layers, as well as the layer outputs, must fit into shared memory \changed{for optimal performance}.
\changed{In contrast, M\"{u}ller~\etAl~\cite{mueller2021realtime} reload the weights from global memory in every layer evaluation per warp.}
As an example, consider a network with $l=4$ layers, each with $c=48$ channels.
Then the weights and biases require $4 * 48^2 * 2$ bytes and $4 * 48 * 2$ bytes, respectively, for a total of $m_w=18816$ bytes that have to be stored once. 
Additionally, each thread stores the $48$ layer outputs, leading to $m_s=48*32*2=3072$ bytes per warp. 
Therefore, with the limitation of $48kB$ shared memory, $w=\lfloor(48k-m_w)/m_s\rfloor=9$ warps can fit into memory. More warps -- up to the hardware limit of 32 -- are advantageous, as they allow to hide pipeline latency by switching between the warps per multiprocessor.
Note that the first (last layer) has to be handled separately, as the input (output) dimension differs.
This results in the maximal network configurations given in \autoref{tab:network-config}.
\begin{table}[tb]
    \caption{Largest possible network configurations for the proposed TensorCore implementation of fV-SRN.}
    \label{tab:network-config}
    \centering
    \begin{tabular}{ | m{5em} | m{0.5cm}| m{0.5cm} | m{0.5cm} | m{0.5cm} | m{0.5cm} | } 
      \hline
      channels & 32 & 48 & 64 & 96 & 128 \\ 
      \hline
      layers & 22 & 10 & 6 & 3 & 2 \\ 
      \hline
    \end{tabular}
\end{table}

The number of Fourier features $m$ (see \autoref{sec:basic}) is chosen as $m=(c-4)/2$, so that the size of the used input vector $\bm{v}$ matches the channel count.
As shown in \autoref{sec:steady:performance}, our TC implementation achieves a speedup of up to $16.8\times$ against 32-bit PyTorch or $9.8\times$ against 16-bit PyTorch.

\subsection{Volumetric Latent Grid}\label{sec:fastSRN:latentgrid}
When using V-SRN with a network configuration that is small enough to enable interactive volume rendering, we observe a significant drop in the networks' prediction skills. The reason lies in the loss of expressive power of the network when relying solely on the few network weights to encode the volume. 
To circumvent this limitation, we borrow an idea proposed by Takikawa~\etAl~\cite{takikawa2021neural} for representing an implicit surface that is encoded as a signed distance function via an SRN. The proposed architecture employs a sparse voxel octree, which stores latent vectors at the nodes instead of distance values. Each octree node stores a trainable $F$-dimensional vector that is interpolated across space and passed as additional input to the SRN network.
Since the SRN learns to predict a single surface, an adaptive voxel octree with a finer resolution near the surface is used.
We adopt this approach of a volumetric latent space but use a dense 3D grid instead of a sparse voxel octree. Especially, since in direct volume rendering it is desirable to change the TF mapping of density values to colors after training, refining adaptively toward a single surface is not suitable.

Let $G:\R^3\rightarrow\R^F$ be a regular 3D grid with $F$ channels, i.e., $F$ parameters per grid vertex, and a resolution of $R$ vertices along each axis. In the interior of each cell, the values are tri-linearly interpolated to obtain a continuous field. 
When evaluating the SRN $V_\Theta$ at position $\bm{x}$, the grid is interpolated at $\bm{x}$ and the resulting latent vector $G(\bm{x})=\bm{z}\in\R^F$ is passed as additional input -- alongside the Fourier features -- to the network.
The contents of $G$ are trained jointly with the network weights and biases.

\begin{figure}
    \centering
    \includegraphics[width=\linewidth]{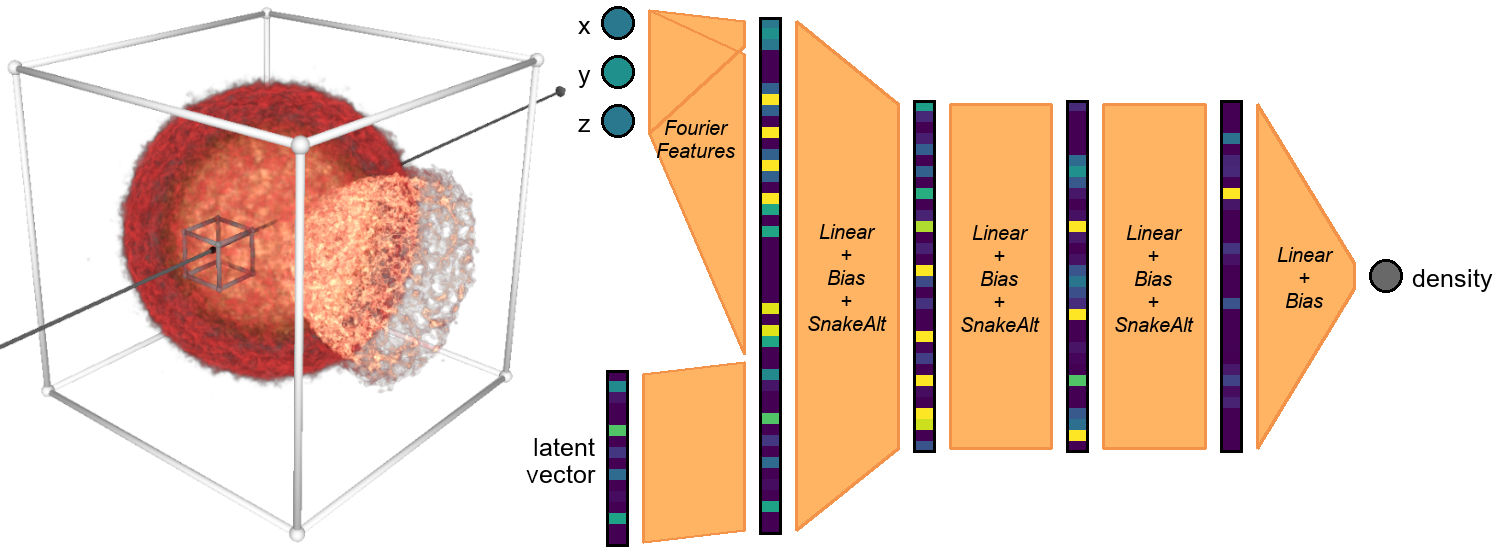}%
    \caption{\changed{Visualization of the proposed fV-SRN architecture: The input position $x,y,z$ is augmented by the Fourier features and the latent vector sampled from the coarse latent grid on the left. The resulting hidden vector with $48$ channels is then passed to an MLP with four layers and $32$ hidden channels to predict the scalar density.}}
    \label{fig:networkvis}
\end{figure}

With this approach, we can keep the network small enough to enable fast inference, up to networks of only two layers {\`a} 32 channels, while maintaining the reconstruction quality of V-SRN. For the evaluation of the network and grid configurations, we refer to \autoref{sec:steady:volumetric}.
We found that the best compromise between speed, quality, and compression rate is achieved with a network of four layers \`a 32 channels and a latent grid of resolution $R=32$ and $F=16$ features.
This configuration is used in \autoref{fig:compressionTeaser}d) and as default in the ablation studies below.   
\changed{The basic network architecture is illustrated in \autoref{fig:networkvis}.}
Compared to \emph{neurcomp}, the proposed network leads to a speedup of up to $9.8\times$ for training and $165\times$ for rendering.
\changed{The latent grid is stored in four 3D CUDA textures with four channels each, so that hardware-supported trilinear texture interpolation can be exploited.}

By using a latent grid, most of the parameters are stored in the grid instead of the network. For the configuration described above,
the grid requires $2$MB of memory (four bytes per voxel and channel), whereas the network consumes only around $7.3$kB of memory.
\changed{Note that in all specifications of the memory consumption of fV-SNR given in this paper, the memory consumed by the latent grid and the network weights are included.
We further avoid storing the latent grid in a float-texture, by using a CUDA feature that enables the use of 8-bit integers per entry that are then linearly mapped to $[0,1]$ in hardware.
Thus, we compute the minimal and maximal grid value for each channel, and use these values to first map the grid values to $[0,1]$ and then uniformly discretize them into 8-bit values.
This reduces the memory footprint of the latent grid representation to a quarter of the size, while reducing the rendering times only slightly by roughly 5$\%$ due to reduced memory bandwidth. At the same time, the quality of the rendered images is slightly decreased by a factor of up to 2$\%$ of the reference SSIM and LPIPS statistics. Visually, however, the discretization does not introduce any perceptual differences, and is used in all of our experiments.} 


\section{Ablation Study}\label{sec:steady}
\begin{figure}[tb]
    \centering
    \begin{overpic}[totalheight=2.5cm]{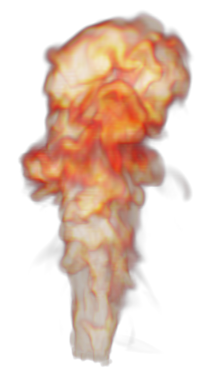}
        \put(0,5){a)}
    \end{overpic}%
    ~~~%
    \begin{overpic}[trim=60 40 20 60,clip,totalheight=2.5cm]{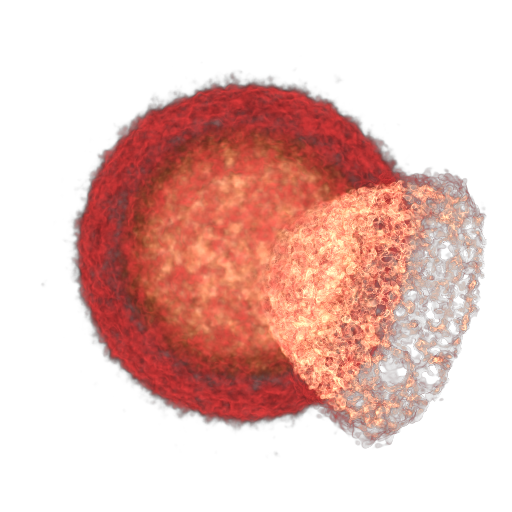}
        \put(0,5){b)}
    \end{overpic}%
    ~~~%
    \begin{overpic}[trim=00 40 20 60,clip,totalheight=2.5cm]{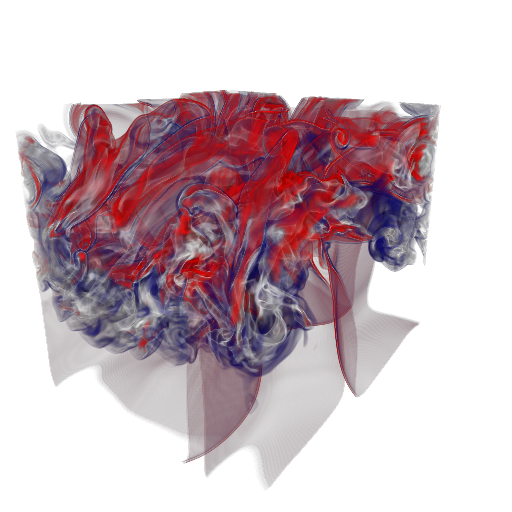}
        \put(0,5){c)}
    \end{overpic}%
    \caption{Datasets used in the ablation study: a) ScalarFlow ($178^3$)~\cite{eckert2019Scalarflow}, b) Ejecta ($256^3$), c) RM  ($256^3$).}
    \label{fig:datasets}
\end{figure}

To select the network architecture with the best reconstruction quality from the possible configurations within the hardware limitations, we trained different networks on three different datasets (see \autoref{fig:datasets}): The ScalarFlow dataset~\cite{eckert2019Scalarflow}---a smoke plume simulation with 500 timesteps, the Ejecta dataset---a supernova simulation with 100 timesteps, and the RM dataset---a Richtmyer-Meshkov simulation with $255$ timesteps. All datasets are given on Cartesian voxel grids, and they are internally represented with 8 Bits per voxel. All timings are obtained on a system running Windows 10, an Intel Xeon CPU with 3.60GHz, and an NVIDIA GeForce RTX 2070.


Unless otherwise noted, we analyze the capabilities of fV-SRN using world-space training on position-density encodings. The networks are training on $256^3$ randomly sampled positions, with a batch size of $128\cdot 64^2$ positions over 200 epochs, an $L_1$ loss function on the predicted outputs, and the Adam optimizer with a learning rate of $0.01$. We use a modified \emph{Snake}~\cite{Liu2020Snake} activation function with enhanced overall slope, i.e., 
\begin{equation} 
SnakeAlt(x)=0.5x+sin^2(x),
\end{equation}
which results in slight improvements of the reconstruction quality, \changed{see supplementary material}. 
After training, the networks are evaluated by rendering $64$ images of resolution $512^2$ from different views of the objects.
The quality of the rendered images is measured using the image statistics SSIM~\cite{wang2004image} and LPIPS~\cite{zhang2018perceptual} using renderings of the initial volumes as references.
\changed{For training from rendered images, we refer to the supplementary material.}

\subsection{Performance Evaluation}\label{sec:steady:performance}

\begin{figure}[tb]
    \centering
    \includegraphics[width=\linewidth]{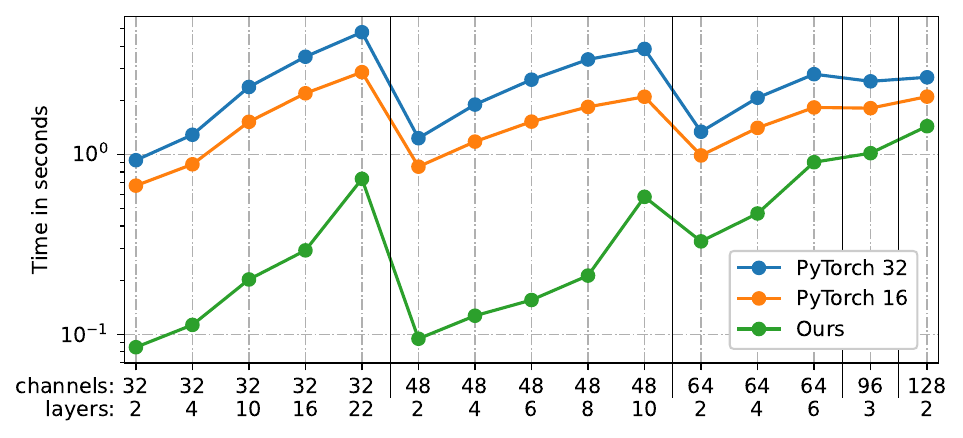}%
    \caption{Performance analysis of fV-SRN using the latent grid and different layer-channel combinations. A $256^3$ dataset is rendered to a $512^2$ viewpoint with a constant stepsize of 1 voxel 
    \changed{without gradients and empty-space skipping}. 
    Measures are obtained as averages over many views. 
    For thin networks with few channels, fV-SRN outperforms a native PyTorch implementation using 32-bit or 16-bit floats by up to a factor of $17\times$. For wide networks with 96 or 128 channels, computation dominates over memory access, leading to a smaller speedup.}
    \label{fig:NetworkConfig-Performance}
\end{figure}

First, we compare the performance of the proposed TC implementation to a native PyTorch implementation of the same architecture. 
Performance measures include the time to access the latent grid and to evaluate fV-SRN with the positional information augmented by Fourier features. 
\autoref{fig:NetworkConfig-Performance} shows the timings for rather lightweight networks as well as 
the largest possible networks within the TC hardware constraints.

As can be seen, the largest speedup of $16.8\times$ ($9.8\times$) over a 32-bit (16-bit) PyTorch implementation is achieved for a medium-sized network of 6 layers and 48 channels. For very small networks of 2 or 4 layers with 32 channels, the speedup goes down to $\approx 11\times$ ($\approx 8\times$). For larger networks, e.g. two layers {\`a} 128 channels, the network evaluation becomes computation bound and the reduction of memory access operations as achieved by our solution becomes less significant. However, also in these cases, a speedup of $1.9$ ($1.5$) against 32-bit (16-bit) PyTorch can still be achieved.
\changed{We notice, however, that fast renderings with 5-10 FPS are only possible with small networks.}

\subsection{Latent Grid}\label{sec:steady:volumetric}

\begin{figure}[t]
    \centering
    \includegraphics[width=\linewidth]{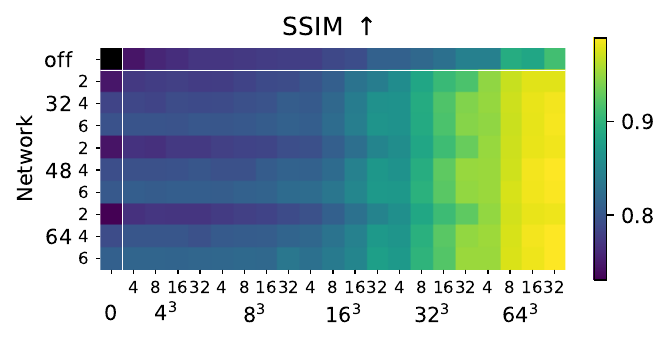}%
    \vspace*{-5pt}%
    \caption{Effect of latent grid resolution on reconstruction quality for Ejecta. The x-axis shows the grid resolution ($4^3,8^3,...,64^3$) and grid feature channels $F$ ($4,8,16,32$), or $0$ if no grid is used. The y-axis represents the network configuration, the number of hidden channels ($32,48,64$), and the number of layers ($2,4,6$). The special network ``off'' shows the results without a network, where only a density grid requiring the same memory as the current latent grid $G$ is used. The LPIPS score and the other datasets show similar behavior.}
    \label{fig:volumetric:stats}
\end{figure}

Next, we investigate the effect of the volumetric latent grid 
on reconstruction quality.
For Ejecta with many fine-scale details, \autoref{fig:volumetric:stats} shows quantitative results for different resolutions of the latent grid and different 
network configurations.
As one can see, the reconstruction quality drastically increases with increasing latent grid resolution. At finer grids, i.e. $32^3$ and higher, the choice of the network has a rather limited effect on the overall reconstruction quality. \changed{The differences between networks of four and six layers are not noticeable}. In these cases, a small network of only four layers {\`a} 32 hidden channels is sufficient to achieve good reconstruction quality. Only when using small grids -- or no latent grid at all -- can larger networks improve the overall quality. 

\begin{figure}[t]
    \centering
    \setlength{\fboxsep}{0pt}
    \setlength{\tabcolsep}{0pt}
    \begin{tabular}{ccc}%
    \begin{overpic}[width=0.32\linewidth,trim=60 40 20 60,clip]{figures/VolumetricFeatures/ejecta70_reference}
        \put(5,5){\fcolorbox{white}{white}{a)}}
    \end{overpic}%
    &%
    \begin{overpic}[width=0.32\linewidth,trim=60 40 20 60,clip]{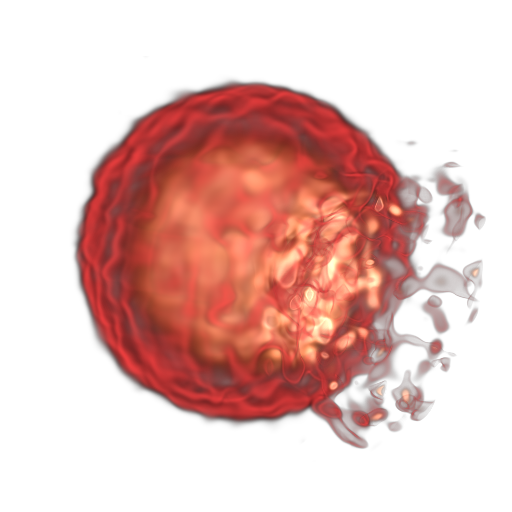}
        \put(5,5){\fcolorbox{white}{white}{b)}}
    \end{overpic}%
    &%
    \begin{overpic}[width=0.32\linewidth,trim=60 40 20 60,clip]{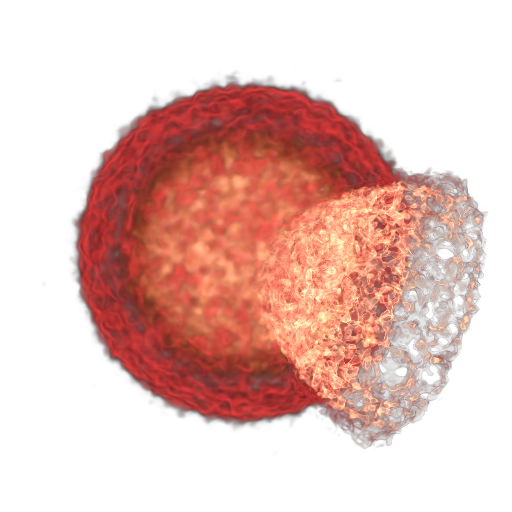}
        \put(5,5){\fcolorbox{white}{white}{c)}}
    \end{overpic}%
    \\[-5pt]%
    &%
    \begin{overpic}[width=0.32\linewidth,trim=60 40 20 60,clip]{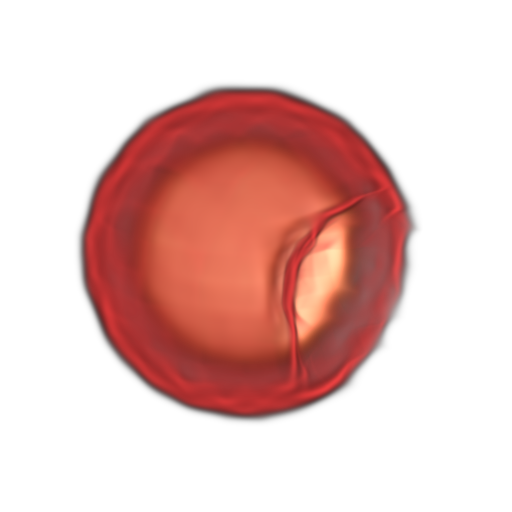}
        \put(5,5){\fcolorbox{white}{white}{d)}}
    \end{overpic}%
    &%
    \begin{overpic}[width=0.32\linewidth,trim=60 40 20 60,clip]{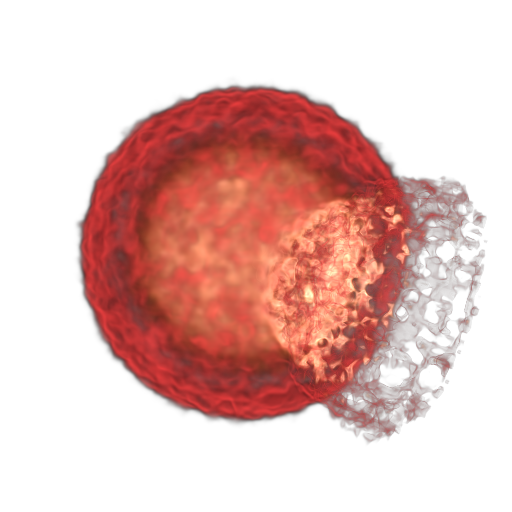}
        \put(5,5){\fcolorbox{white}{white}{e)}}
    \end{overpic}%
    \end{tabular}%
    \caption{Effect of latent grid resolution on reconstruction quality. a) Reference. b) Latent grid of $R=8, F=8$ with a network of 32 channels, 2 layers. c) Latent grid of $R=32, F=16$ with a network of 32 channels, 2 layers. d) and e) Density grid without a network of equivalent memory consumption as b) and c).}
    \label{fig:volumetric:images}
\end{figure}

To confirm that the quality improvement is not solely due to the volumetric latent grid while the SRN is superfluous, we compare the rendered images to images that were rendered from a low-pass filtered density grid with the same memory consumption as the latent grid (row "off" in \autoref{fig:volumetric:stats}). For example, for a latent grid of resolution $32^3$ and $16$ features, the original volume is first low-pass filtered and then down-sampled to a grid resolution of $\sqrt[3]{32^3 \cdot 16}\approx 81$. The width of the low-pass filter is selected according to the sub-sampling frequency. 
As one can see from \autoref{fig:volumetric:stats}, and evidenced by the qualitative assessment in \autoref{fig:volumetric:images}, by using a latent grid in combination with the SRN even small-scale structures are maintained. In the low-resolution density grid, many of these structures are lost. 

Notably, since the latent grid can be trained very efficiently and takes the burden from the SRN to train a huge number of parameters, the training times of fV-SRN are up to a factor of $9.8\times$ faster than those of V-SRN with the same total number of parameters, see \autoref{fig:compressionTeaser}. Especially because SRNs overfit to a certain dataset and training has to be repeated for each new dataset, we believe that this reduction of the training times is mandatory to make SRNs applicable.  
%

\subsection{Fourier Features}\label{sec:steady:fourierimportance}

\begin{figure}
    \centering
    \includegraphics[width=\linewidth]{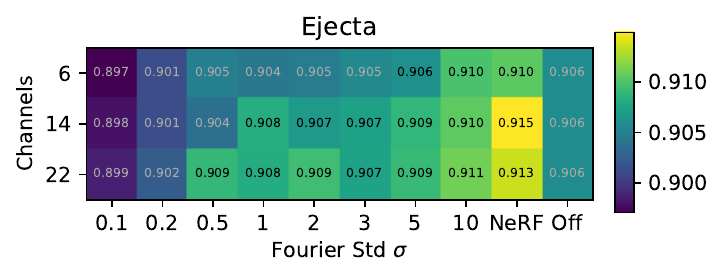}%
    \vspace*{-3pt}%
    \caption{SSIM values for different types of Fourier features -- Tancik~\etAl{} with different values for $\sigma$, NeRF, and disabled features -- and different feature sizes $m$, evaluated on the Ejecta dataset. The LPIPS score and the other datasets show similar behavior.}
    \label{fig:fouriergrid}
\end{figure}

In the following, we shed light on the effects of Fourier features on the overall reconstruction quality of networks that were trained in world-space for density prediction.
We compare the construction of Fourier features according to Mildenhall~\etAl~\cite{mildenhall2020nerf}, denoted ``NeRF'', and Tancik~\etAl~\cite{tancik2020fourier} with standard deviation $\sigma$ as hyperparameter. In addition, we evaluate the reconstruction quality when Fourier features are not used. Networks were trained for multiple values of $\sigma$ and three different numbers of Fourier features. The results can be found in \autoref{fig:fouriergrid}. They demonstrate the general improvements due to the use of Fourier features, and furthermore indicate the superiority of ``NeRF'' over random Fourier feature by Tancik~\etAl{} in combination with our network design.  


\subsection{Density vs. Color Prediction}\label{sec:steady:densitycolor}
Next, we shed light on the reconstruction quality for networks that predict densities that are then mapped to colors via a user-defined TF (the approach we have followed so far), and networks that directly predict colors at a certain domain location. In the latter case, the network encodes colors dependent on positions, and the loss function considers the differences between the encoded colors and the colors that are obtained by post-shading at the interpolated input positions. 

Density prediction enables to change the TF after a sample has been reconstructed \changed{without retraining}. When predicting colors, however, the network needs to be re-trained whenever the TF is changed. Hence, this approach seems to be less useful in practice, yet it is interesting to analyze how well a network can adapt its learning skills to those regions emphasized by a TF mapping. Possibly, the network can learn to spend its capacities on those regions that are actually visible after the TF has been applied, which may result in improved reconstruction quality. 

\begin{figure}
    \centering
    \begin{overpic}[width=0.24\linewidth,trim=0 20 0 0,clip]{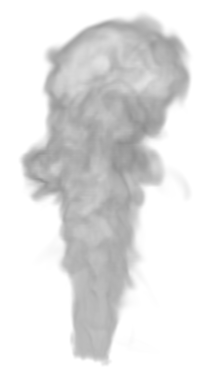}
        \put(0,0){\includegraphics[width=0.2\linewidth,cfbox=gray 1pt 0pt]{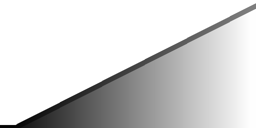}}
    \end{overpic}%
    ~~
    \begin{overpic}[width=0.24\linewidth,trim=0 20 0 0,clip]{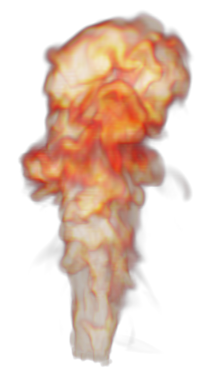}
        \put(0,0){\includegraphics[width=0.2\linewidth,cfbox=gray 1pt 0pt]{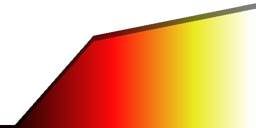}}
    \end{overpic}%
    ~~
    \begin{overpic}[width=0.24\linewidth,trim=0 20 0 0,clip]{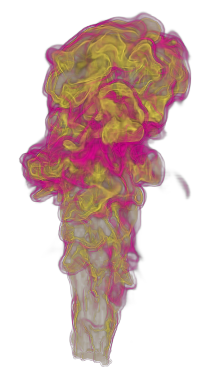}
        \put(0,0){\includegraphics[width=0.2\linewidth,cfbox=gray 1pt 0pt]{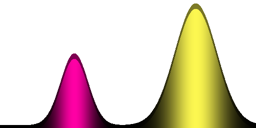}}
    \end{overpic}%
    \\%
    \includegraphics[width=\linewidth]{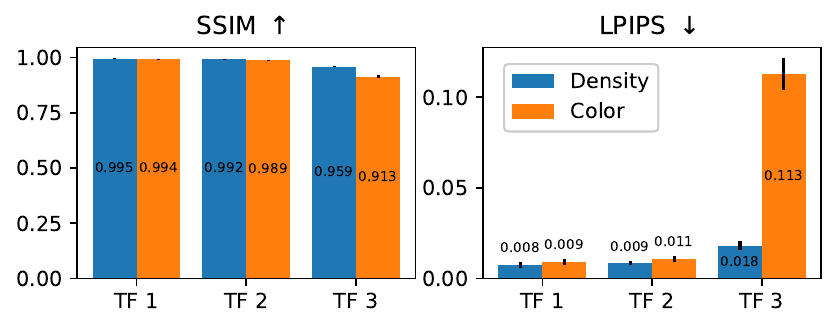}%
    \caption{Quality comparison of fV-SNR-based density and color prediction. 
    The tests were conducted with three different TFs, reference images \changed{and transfer functions} are shown in the first row.}
    \label{fig:DensityVsColor}
\end{figure}

\changed{%
We trained four instances of fV-SRN: One that predicts densities and three networks that predict colors that have been generated via three different TFs on ScalarFlow. Reference images for the three TFs are shown in \autoref{fig:DensityVsColor}, combined with the achieved reconstruction quality.
%
For the first two TFs, there are almost no differences in image quality between density and color prediction. For the third TF with two narrow peaks, however, color prediction performs considerably worse, even though the network needs to learn significantly fewer positions at which a non-transparent color is assigned. We hypothesize that especially narrow peaks in the TF make the prediction difficult. In such cases, the absorption changes rapidly over a short interval, so that the network training on uniformly distributed locations cannot adequately learn these high frequencies.
}

To force the network to consider more positional samples in regions where the TF mapping generates a color, we propose the following adaptive resampling scheme:
After each $N$-th epoch ($N=50$ in our experiments), we evaluate the prediction error over a coarse voxel grid $E$ of resolution $128^3$. Per voxel, $8$ positions are sampled and evaluated as an approximation of the average prediction error per voxel.
$E$ is then used to sample new training data for the next $N$ epochs, where the number of sampled positions is made proportional to the values in $E$. This allows the training process to focus on regions with high prediction error and samples the volumetric field in more detail in these regions.

By using the proposed adaptive sampling scheme, e.g., the LPIPS score for TF~3 for the color-predicting network is improved from $0.113$ to $0.039$. Even though, however, we do not believe that the quality of color prediction can match the quality of density prediction when rather sharp TFs are used. Thus, and also due to the restriction of color prediction to a specific TF, we consider this option to be useful only when a color volume is given initially.

\section{Gradient Prediction and Higher-Order Derivatives}\label{sec:gradients}
\changed{%
In the following, we shed light on the use of fV-SRN to learn a mapping that not only predicts a scalar field value at a given position but also the gradient and even higher order derivatives at that position. 
The gradient is important in volume rendering to apply gradient-magnitude-based opacity and color selection via TF mappings ~\cite{levoy1988display}, 
and, since the gradient at a certain position is the normal vector of the isosurface passing through this position, to illuminate the point, e.g., via Phong lighting.
}

\begin{figure}
    \centering
    \includegraphics[width=\linewidth]{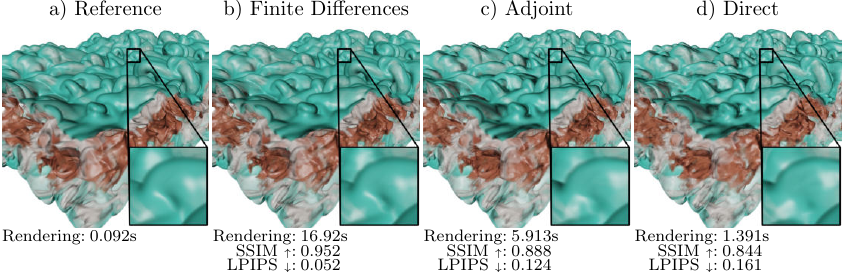}%
    \caption{\changed{Network-based direct volume rendering including gradient-magnitude-based shading. a) Gradients are computed via finite differences, i.e. calling the network 8 times for gradient estimation. b) Gradients are computed via the adjoint method, by backpropagation through the volume, c) fV-SRN is trained to predict scalar values and gradients. Shown is Jet with a compression ratio of $1:985$ and Phong shading. fV-SRN without gradient estimation (i.e., no shading) requires $1.373$ seconds.}}
    \label{fig:gradientteaser}
\end{figure}

\changed{
In particular, we evaluate different strategies to estimate gradients in network-based scalar field reconstruction: Using finite differences by calling the network multiple times (\emph{FD}), using the adjoint method (\emph{Adjoint}), and training fV-SRN to predict the gradients alongside the density. 
As we will show, the latter improves the rendering performance by $\approx 12\times$ over finite differences and $4-5\times$ over the adjoint method, while reducing the quality only slightly, see \autoref{fig:gradientteaser}.
All three methods are implemented using the proposed TC kernel (\autoref{sec:fastSRN:TC}).
}

\changed{%
The common method to compute gradients during volume ray-casting is to use \emph{FD}, more concretely central differences, between trilinearly interpolated scalar values with a step size of one voxel size ~\cite{levoy1988display}. Compared to computing analytical derivatives of the trilinear interpolant, the use of central differences avoids discontinuities at the voxel borders.
Since the use of \emph{FD} (with fV-SRN) introduces a bias if the same method (w/o network) is used as reference, \emph{FD} with fV-SRN leads to the best prediction in general, see \autoref{fig:gradientteaser}b.
This method, however, introduces a large computation overhead as seven network evaluations are required to compute the density and gradients.
Therefore, in previous work the adjoint method (\emph{Adjoint}) was proposed as an alternative ~\cite{davies2020effectiveness,lu2021compressive}. \emph{Adjoint} uses backpropagation through the trained reconstruction network to predict the change of the scalar value depending on changes of the position, see \autoref{fig:gradientteaser}c.
}

\changed{%
The fastest prediction is achieved by extending the output of fV-SRN to predict scalar values and gradients, see \autoref{fig:gradientteaser}d.
However, as the network now needs to predict four outputs -- density plus gradient $x,y,z$ -- instead of one within the same network weights and latent grid size, the quality is slightly reduced.
For an extended comparison on further datasets, including implicit functions with analytical gradients, and a detailed study on how to design the loss function to include the gradients, we refer to the supplementary material.
}

\begin{figure}%
	\centering
	\begin{overpic}[width=0.21\linewidth]{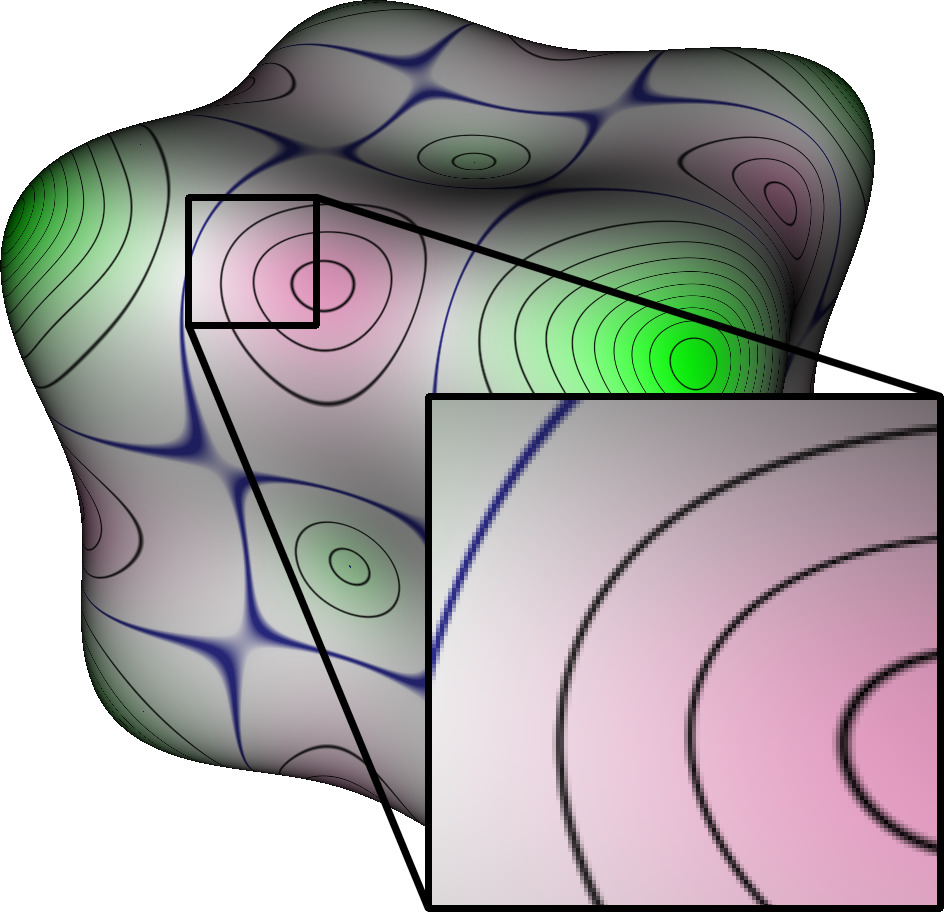}
		\put(5,5){a)}%
	\end{overpic}%
	\begin{overpic}[width=0.21\linewidth]{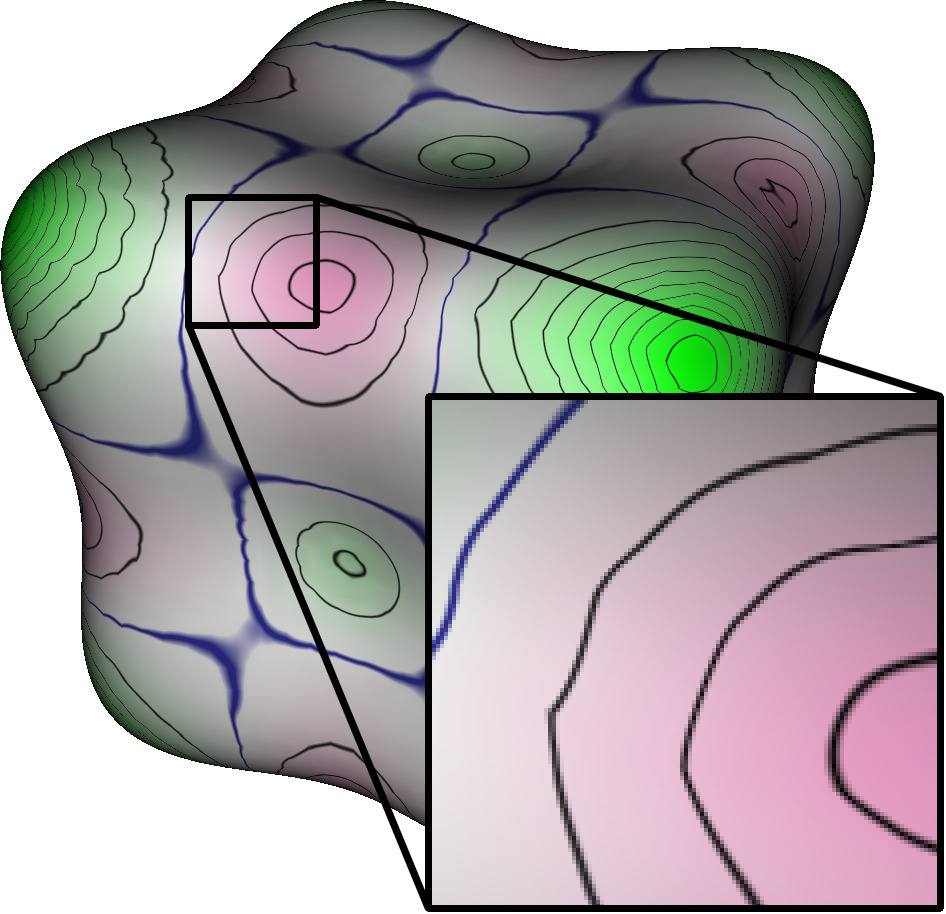}
		\put(5,5){b)}%
	\end{overpic}~~%
	\begin{overpic}[width=0.24\linewidth]{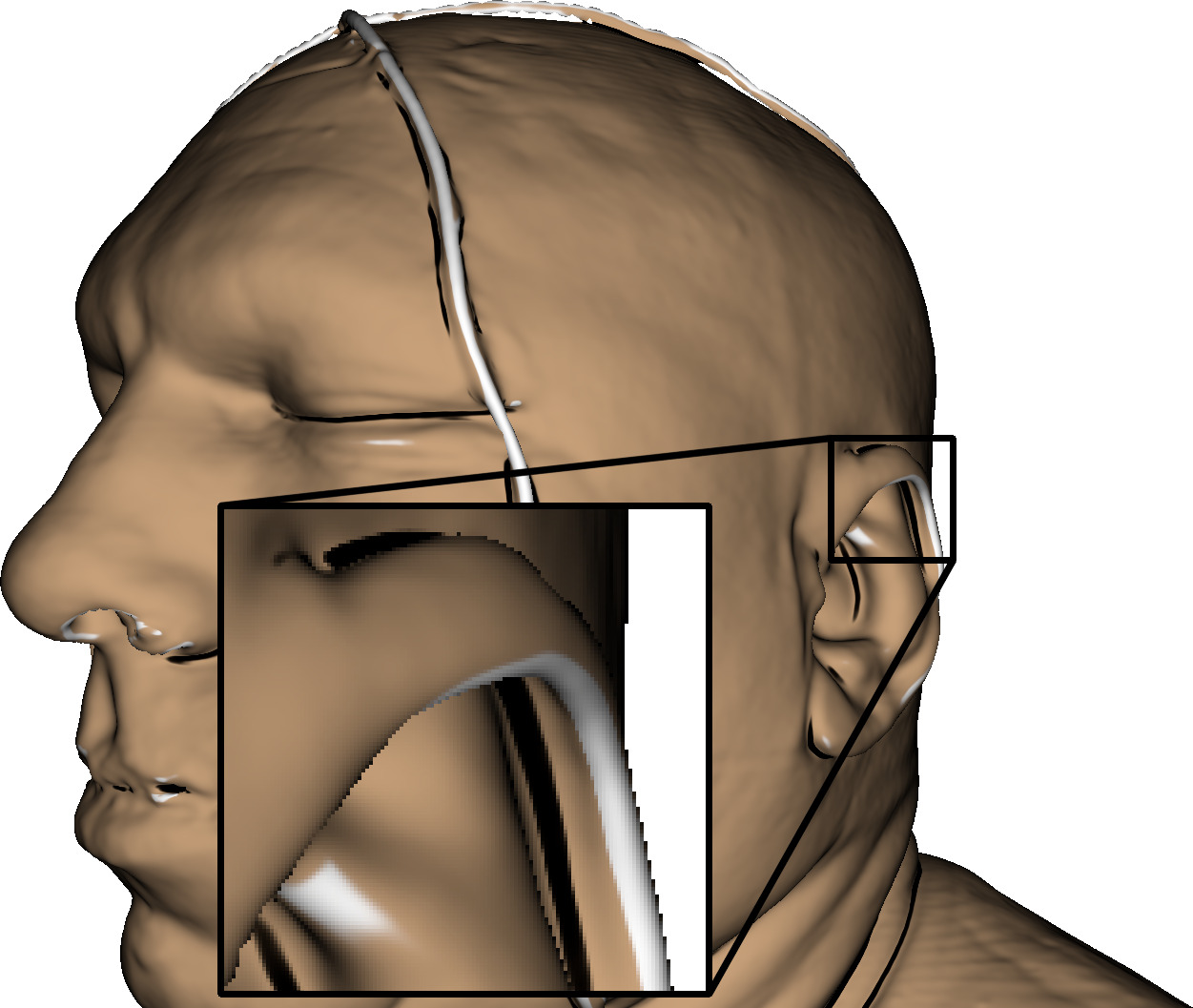}
		\put(-5,5){c)}%
	\end{overpic}~~%
	\begin{overpic}[width=0.24\linewidth]{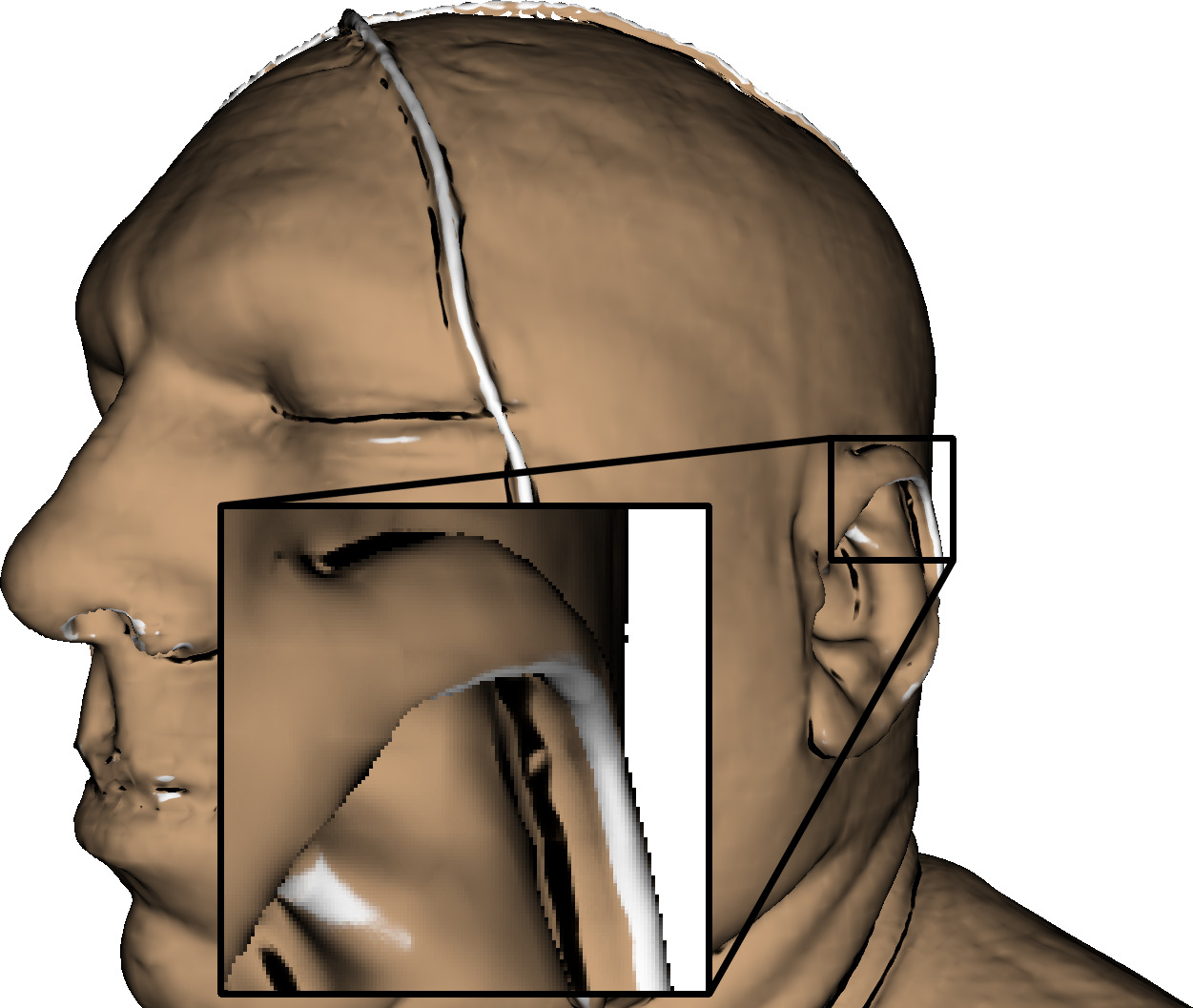}
		\put(-5,5){d)}%
	\end{overpic}%
	\caption{\changed{fV-SRN for isosurface rendering with curvature-based TF mapping. (a,b) Gaussian curvature $\kappa_1\kappa_2$ on an implicit dataset~\cite{kindlmann2003curvature}, (c,d) principal curvatures $\kappa_1$ and $\kappa_2$ are mapped to color on an isosurface in a CT scan. (a,c) are the references, (b,d) show the predictions by fV-SRN.}}%
	\label{fig:curvature}%
\end{figure}

\changed{%
If higher-order derivatives are required, e.g., for TFs incorporating curvature measures~\cite{kindlmann2003curvature}, \emph{FD} and \emph{Adjoint} become increasingly intractable.
For finite differences, Kindlmann~\etAl~\cite{kindlmann2003curvature} propose a stencil with a support of $4^3$ samples. This would require $64$ network evaluations per sample along the ray. Similarly, the adjoint method requires an additional adjoint pass per row in the Hessian matrix.
As an outlook for future research, we show that the SRNs can be trained to jointly predict densities, gradients, and also curvature estimates as a multi-valued output. 
First results using the shading proposed by Kindlmann~\etAl~\cite{kindlmann2003curvature} on isosurface renderings are given in \autoref{fig:curvature}.
}

\section{Performance and Quality Comparison}\label{sec:baselines}
\changed{
In the following, the quality and performance of fV-SRN is compared to
\emph{neurcomp}~\cite{lu2021compressive}, TThresh~\cite{ballester2019tthresh}, and cudaCompress~\cite{treib2012interactive,treib2012turbulence}. We compare to TTresh because of the extreme compression rates it can achieve, and to cudaCompress because of its decoding efficiency.
The publicly available implementations of TThresh (running on the CPU) and cudaCompress (running on the GPU) are used.
For the comparison, we chose Jet with Phong shading, as introduced in \autoref{sec:gradients}, and Ejecta at a resolution of $1024^3$, see \autoref{fig:compressionTeaser}.
To achieve a given compression ratio, fV-SRN changes the latent grid resolution, \emph{neurcomp} adapts the number of hidden channels, TTresh modifies the bitplane cutoff, and cudaCompress adapts the stepsize for quantizing discrete wavelet coefficients. Further results on additional datasets are given in the supplementary material.
}


\begin{figure*}
    \centering
    \includegraphics[width=\linewidth]{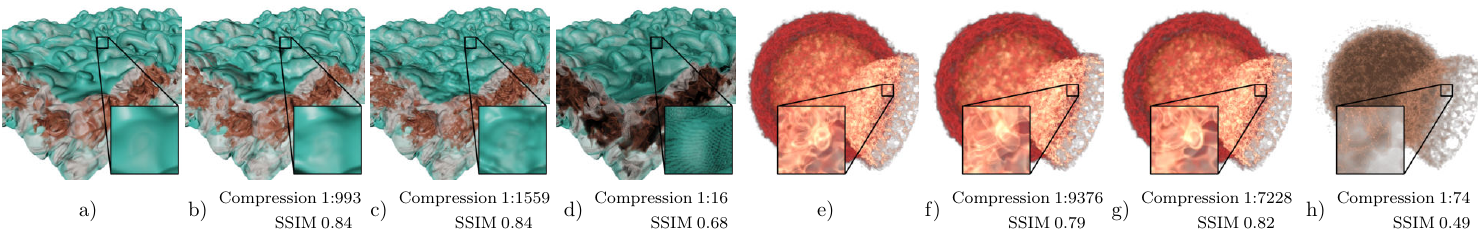}%
	\caption{
	\changed{Visual comparison of volume compression schemes for Jet (a-d) and Ejecta (e-h). (a,e) Reference, (b,f) fV-SRN, (c,g) TThresh, (d-h) cudaCompress. For each scheme, the result obtained with a compression ratio closest to 1:1000 and 1:10,000 for Jet and Ejecta, respectively, is selected from \autoref{fig:CompressionExtended:Plot}.}}
    \label{fig:CompressionExtended:Images}
    \vspace*{5pt}
    \includegraphics[width=\textwidth]{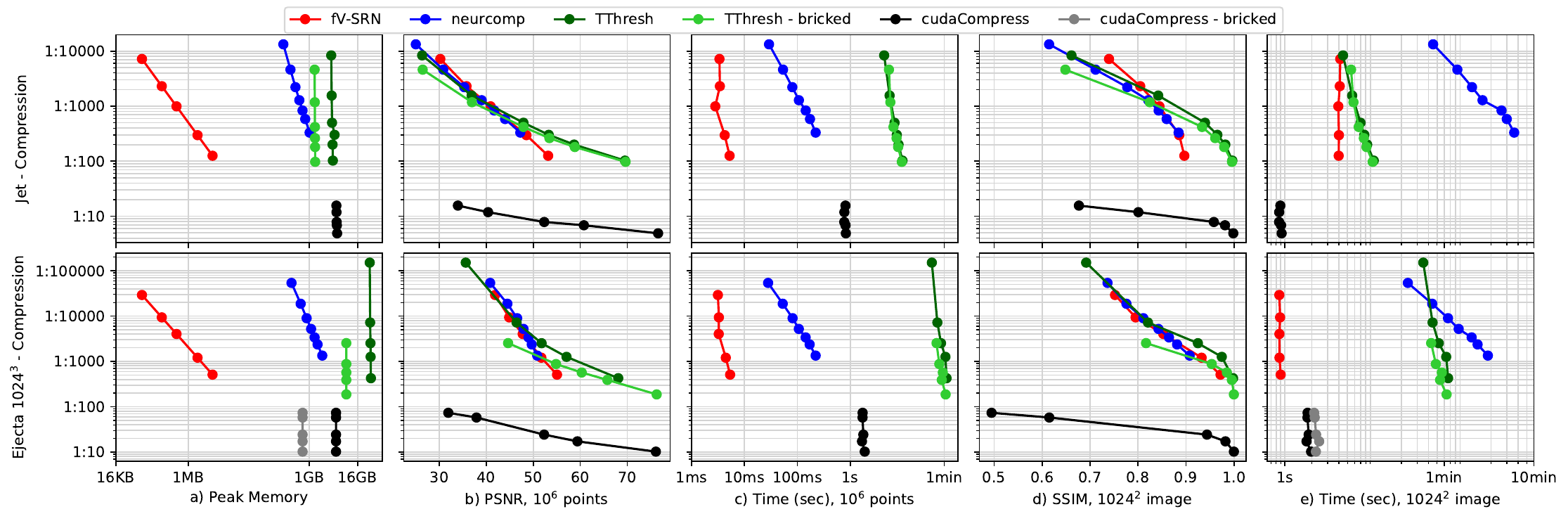}%
    \vspace*{-5pt}%
    \caption{\changed{Evaluation of different compression rates of fV-SRN, \emph{neurcomp}~\cite{lu2021compressive}, TThresh~\cite{ballester2019tthresh}, and cudaCompress~\cite{treib2012interactive,treib2012turbulence}, using Jet of resolution $512\times 336\times 768$ and Ejecta of resolution $1024^3$. Jet uses Phong shading, the adjoint method is used by the network-based approaches to predict the gradients, see \autoref{sec:gradients}.}
    }
    \label{fig:CompressionExtended:Plot}
\end{figure*}

\changed{%
For a quantitative evaluation, compression ratios of the four methods are plotted against a) the peak CPU and GPU memory required for decoding, b,c) the time it requires to reconstruct $10^6$ random locations as well as the resulting PSNR, d,e) the time it requires to render an image of resolution $1024^2$ with 2 samples per voxel on average as well as the resulting SSIM statistics, see \autoref{fig:CompressionExtended:Plot}.
All timing statistics are performed on an Intel Xeon CPU with 8 cores and 3.60GHz, equiped with a NVIDIA GeForce RTX 2070 GPU.
}

\changed{%
Regarding PSNR and SSIM, fV-SRN, \emph{neurcomp} and TThresh are almost on-par. For high compression ratios, the network-based approaches slightly outperform TThresh, while the opposite is true at low compression ratios.
However, both TThresh and cudaCompress require additional temporal memory, as they need to decode the volume before rendering. For TThresh and cudaCompress, respectively, the temporarily required memory can grow up to $34$GB and $4.7$GB.
\emph{neurcomp} requires temporal memory to store the hidden states during network evaluation, computed here for evaluating $1024^2$ rays in parallel.
As shown in \autoref{sec:fastSRN:TC}, fV-SRN runs completely in shared memory and requires no additional temporal memory for evaluation, besides storing the latent-space representation including network weights and latent grid -- for sampling and rendering.
}

\changed{
Treib~\etAl~\cite{treib2012interactive,treib2012turbulence} propose bricked decompression and rendering in combination with cudaCompress. In our case, we use a brick size of $256^3$. This drastically reduces the memory requirements from $4.7$GB to around $700$MB, while increasing the rendering time by roughly $13\%$ for the Ejecta dataset. Note that this bricked rendering is only possible for the regular access pattern during rendering. For random access, the whole volume still needs to be decompressed.
In a similar fashion, we applied a bricked TThresh, where each brick is compressed independently. For Ejecta, this also reduces the memory requirement from $34$GB to around $8$GB, but also drastically reduces the achieved compression ratio.
}

\changed{%
A qualitative comparison of the errors introduced by all compression schemes is given in \autoref{fig:CompressionExtended:Images}.
cudaCompress quantizes the values which introduce large errors with narrow TFs. TThresh introduces slight grid artifacts, and both fV-SRN and \emph{neurcomp} blur the dataset at higher compression ratios.
}

%

\section{fV-SRN for Temporal Super-Resolution}\label{sec:unsteady}

We now analyze the extension of fV-SRNs to interpolate between different instances in time of a scalar field. The interpolation should smoothly transition between the instances to create plausible intermediate fields, and topological changes should be handled.
The proposed approach is inspired by previous works by Park~\etAl~\cite{park2019deepsdf} and Chen and Zhang~\cite{chen2019learning}, where latent vectors representing different objects are interpolated to morph one object into another one in a feature-preserving manner. To achieve the aforementioned goals, we extend the volumetric latent spaces, see \autoref{sec:steady:volumetric}, to include the time domain. 
%

Let $\mathfrak{T}_\text{all}=\{1,2,...,T\}$ be the indices of the $T$ timesteps that are available in the dataset.
To save memory, the volumetric latent space is provided only at certain timesteps that we call \emph{keyframes}.
Let $\mathfrak{T}_\text{key}\subset\mathfrak{T}_\text{all}$ be the timestep indices of the keyframes and the volumetric latent space is then indexed as $G_t:R^3\rightarrow\R^F, t\in\mathfrak{T}_\text{key}$.
For timesteps that are between two keyframes, the volumetric latent space is linearly interpolated in time and passed to the network.
During training, timesteps from $\mathfrak{T}_\text{train}\subset\mathfrak{T}_\text{all}$ are used.

\begin{figure*}[p]
    \centering%
    \includegraphics[width=0.97\textwidth]{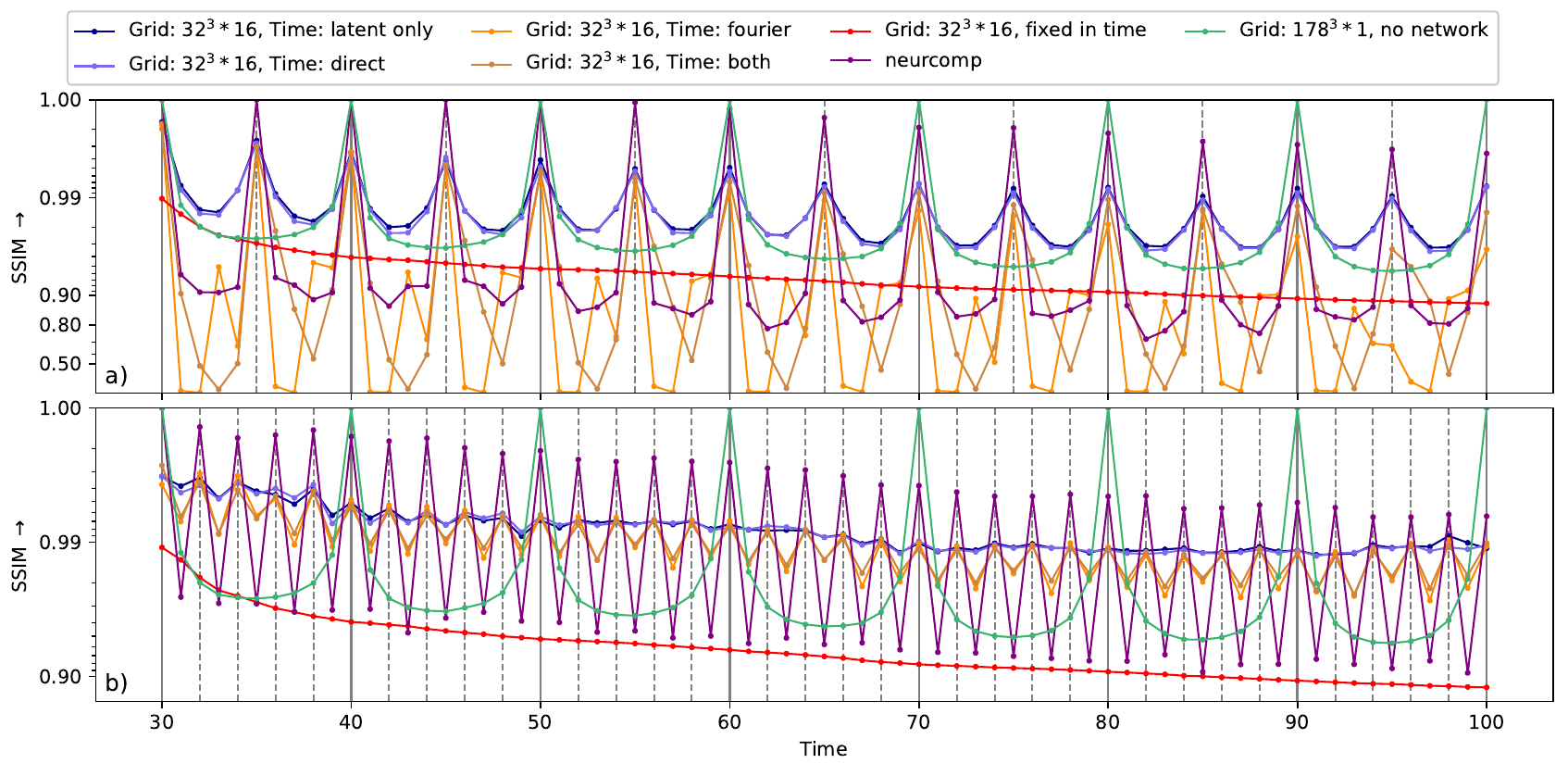}%
    \caption{Analysis of temporal super-resolution of ScalarFlow, showing reconstruction quality for different grid sizes and time encoding schemes. The networks were trained on every 5\textsuperscript{th} timestep in (a) and every 2\textsuperscript{nd} timestep in (b), with a keyframe at every 10\textsuperscript{th} timestep, and evaluated on all timesteps. The blue and yellow lines represent the the different time encoding schemes, green the baseline where the original volume at the keyframes is interpolated in time, and red is the method inspired by Lu~\etAl~\cite{lu2021compressive}, where only a single grid is used and the time is sorely interpreted by the network. \changed{The unchanged neurcomp architecture by Lu~\etAl{}, where all information is stored in the network weights, is depicted in purple.}}
    \label{fig:time:quantitative}
%
    \vspace*{5pt}
    \includegraphics[width=\textwidth]{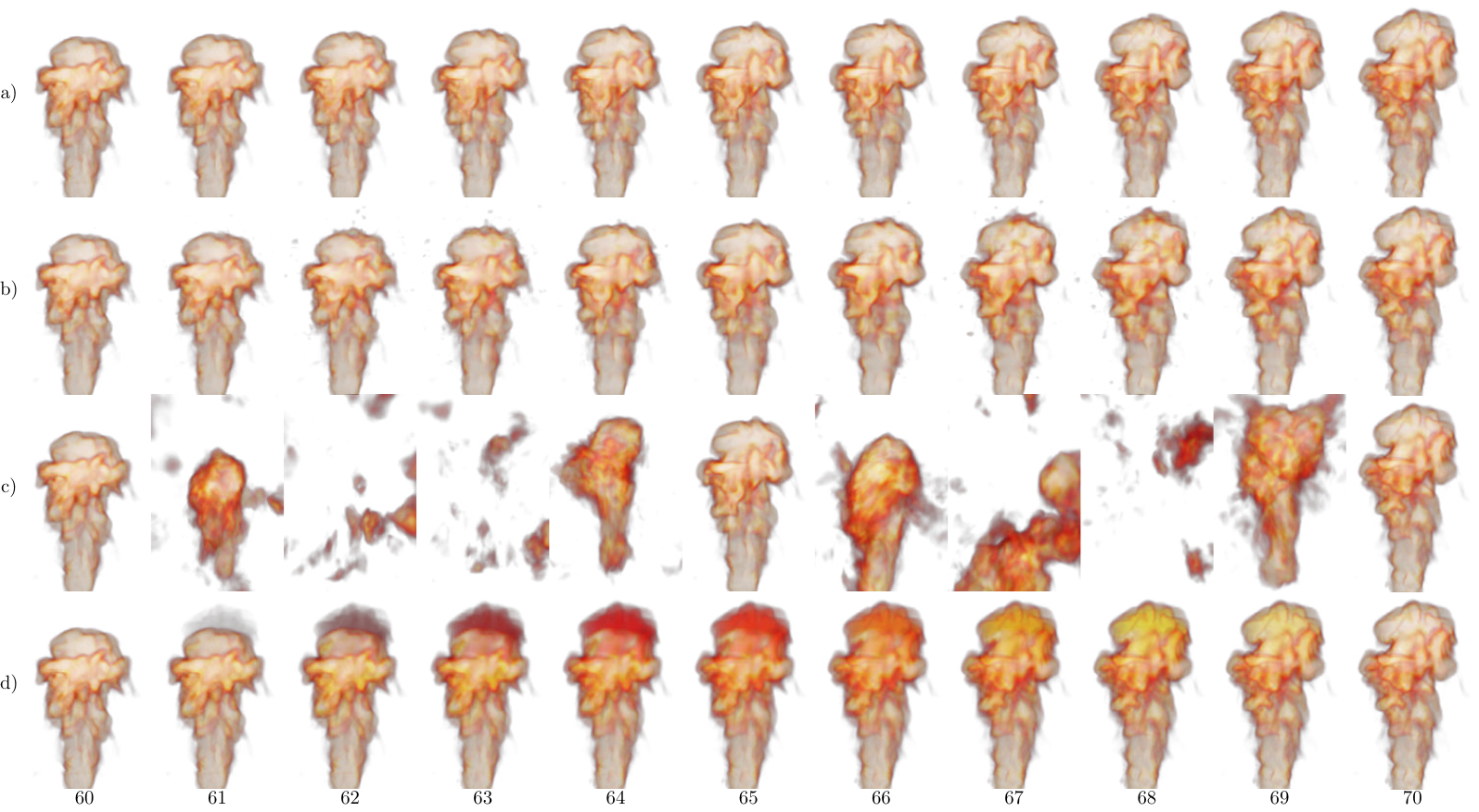}%
    \caption{Qualitative results for timesteps 60 to 70 of ScalarFlow, shown in \autoref{fig:time:quantitative}a. a) Reference, b) Best time encoding ``direct'', \changed{c) ``neurcomp''}, d) Linear interpolation on the initial grids.}
    \label{fig:time:qualitative}
\end{figure*}

In addition to the time-dependent latent space, we evaluate four options to encode the time dimension in the network, so that plausible interpolation is achieved: no extra input (``latent only''); time as an additional scalar input (``direct''); time modulated by Fourier features based on Mildenhall~\etAl{} with $L=4$, see \autoref{sec:steady:fourierimportance} (``fourier''); time as scalar input and Fourier features (``both'').
Quantitative results are given in \autoref{fig:time:quantitative} on the ScalarFlow dataset with a keyframe every 10\textsuperscript{th} timestep for timesteps 30 to 100.
For training, every 5\textsuperscript{th} timestep (\autoref{fig:time:quantitative}a) or every 2\textsuperscript{nd} timestep (\autoref{fig:time:quantitative}b) was used.
For timesteps 60 to 70, \autoref{fig:time:qualitative} shows the qualitative results.

We found that ``latent only'' and ``direct'' lead to good generalization for in-between timesteps that were never seen during training, with no noticeable difference between both methods (\autoref{fig:time:quantitative} blue, \autoref{fig:time:qualitative}b). Those two architectures lead to a  semantically plausible interpolation, that becomes especially noticeable when compared against a baseline (\autoref{fig:time:quantitative} green, \autoref{fig:time:qualitative}d) where the original grid is used at the keyframes and then linearly interpolated in time.

The options including Fourier features in the time domain (``fourier'' and ``both''), however, show chaotic behavior for in-between timesteps (\autoref{fig:time:quantitative} yellow). As opposed to Fourier features in the spatial domain where all fractional positions could have been observed due to the random sampling of the positions, in the time domain only a discrete subset of timesteps are seen. Therefore, during generalization, the Fourier encoding produces value ranges for the network that were never seen before.

Let us also emphasize that \changed{neurcomp} by Lu~\etAl~\cite{lu2021compressive} also supports super-resolution in the time domain, by sending the time domain directly as input to the network, \changed{see \autoref{fig:time:quantitative} purple}. \changed{Neurcomp allows an accurate prediction of the timesteps from the training datasets, using the same compression ratio as fV-SRN, but fails to generalize to in-between timesteps. This can also be clearly seen in the qualitative comparison \autoref{fig:time:qualitative}c. We hypothesize that the time-interpolated latent grid acts as a regularizer in that regard.}
\changed{The importance of a time-varying latent grid is also supported by the following test, \autoref{fig:time:quantitative} red. Using the time encoding ``direct'', but with only a single keyframe for the grid, leads to inferior results.}

\changed{%
In total, fV-SRN allows for an efficient and plausible interpolation in time.
The training time when including the time domain, however, increases drastically. Training a network on every 5\textsuperscript{th} timestep requires around 3:45h.
Using every 2\textsuperscript{nd} timestep instead of every 5\textsuperscript{th} improves the quality of the interpolation (\autoref{fig:time:quantitative}b versus a), but the training time increases accordingly to almost $8$ hours. 
}



\section{Conclusion}
We have analyzed SRNs for compression domain volume rendering, and introduced fV-SRN as a novel extension to achieve significantly accelerated reconstruction performance. Accelerated training as well as the adaptation of fV-SRN to facilitate temporal super-resolution have been proposed.
As key findings we see that
\begin{itemize}
    \item by using custom evaluation kernels and a latent grid, SRNs have the potential to be used in interactive volume rendering applications,
    \item \changed{fV-SRN is an alternative to existing volume compression schemes at comparable quality and significantly improved decoding speed, or similar performance but significantly higher compression ratios,}
    \item SRNs using latent space interpolation can preserve features that are lost using traditional interpolation and enable temporal super-resolution at arbitrary temporal resolution.
\end{itemize}



\changed{In the context of volume rendering, it will be important to investigate the capabilities of SRNs to learn mappings that consider a view-dependent level-of-detail (LoD). In particular, the network might be able to infer more than just values of a low-pass filtered signal, but infer values as they would be perceived when looking at the data through a pixel and perform area-weighted super-sampling. Such a view-dependent learning of LoDs can avoid missing details which are smoothed out using the classical low-pass filtering approach.}

We see further potential in SRNs for scientific data visualization due to their ability to randomly access samples from the compressed feature representation. Due to this property, we see a promising application in the context of flow visualization. By using SRNs to encode position-velocity relationships, particle tracing \changed{or streamline tracing}, with its sparse and highly irregular data access patterns, can work on a compactly encoded vector field representation.


As another interesting use case for fV-SRN we see ensemble visualization. In particular, we intend to investigate whether the idea of multiple latent grids introduced for time-dependent fields can be used to represent similar and dissimilar parts in each ensemble member. An interesting experiment will be to generate Mean-SRNs, which are trained using position-density encodings corresponding to different datasets. This may also give rise to alternative ensemble compression schemes, where differences to a reference are encoded. Furthermore, we plan to investigate whether time and ensemble information can be decoupled in the latent grid. This can eventually enable to retrain ensemble features for a novel ensemble member, and vary the temporal features to predict the temporal evolution.
Finally, we note that including the time domain vastly increases the training time. Thus, similar to the adaptive spatial sampling presented in this work, we plan to investigate adaptive (re-)sampling strategies in time, to focus only on those timesteps that exhibit the largest prediction errors.

\bibliographystyle{eg-alpha-doi}

\bibliography{main}

\newcommand{\etalchar}[1]{$^{#1}$}
\begin{thebibliography}{\uppercase{PCPMMN21}}

\bibitem[BHP14]{10.2312:eurovisstar.20141175}
\textsc{Beyer J., Hadwiger M., Pfister H.}:
\newblock {A Survey of GPU-Based Large-Scale Volume Visualization}.
\newblock In \emph{EuroVis - STARs} (2014), Borgo R., Maciejewski R., Viola I.,
  (Eds.), The Eurographics Association.

\bibitem[BLL19]{berger2018generative}
\textsc{Berger M., Li J., Levine J.~A.}:
\newblock A generative model for volume rendering.
\newblock \emph{IEEE transactions on visualization and computer graphics 25}, 4
  (2019), 1636--1650.

\bibitem[BMT{\etalchar{*}}21]{barron2021mip}
\textsc{Barron J.~T., Mildenhall B., Tancik M., Hedman P., Martin-Brualla R.,
  Srinivasan P.~P.}:
\newblock {Mip-NeRF}: A multiscale representation for anti-aliasing neural
  radiance fields.
\newblock \emph{arXiv preprint} (2021).
\newblock \href {http://arxiv.org/abs/2103.13415} {\path{arXiv:2103.13415}}.

\bibitem[BRLP19]{ballester2019tthresh}
\textsc{Ballester-Ripoll R., Lindstrom P., Pajarola R.}:
\newblock {TTHRESH}: Tensor compression for multidimensional visual data.
\newblock \emph{IEEE transactions on visualization and computer graphics 26}, 9
  (2019), 2891--2903.

\bibitem[CLI{\etalchar{*}}20]{chabra2020deep}
\textsc{Chabra R., Lenssen J.~E., Ilg E., Schmidt T., Straub J., Lovegrove S.,
  Newcombe R.}:
\newblock Deep local shapes: Learning local sdf priors for detailed 3d
  reconstruction.
\newblock In \emph{European Conference on Computer Vision} (2020), Springer,
  pp.~608--625.

\bibitem[CZ19]{chen2019learning}
\textsc{Chen Z., Zhang H.}:
\newblock Learning implicit fields for generative shape modeling.
\newblock In \emph{Proceedings of the IEEE/CVF Conference on Computer Vision
  and Pattern Recognition} (2019), pp.~5939--5948.

\bibitem[DC16]{di2016fast}
\textsc{Di S., Cappello F.}:
\newblock Fast error-bounded lossy hpc data compression with {SZ}.
\newblock In \emph{2016 IEEE international parallel and distributed processing
  symposium (IPDPS)} (2016), IEEE, pp.~730--739.

\bibitem[DMG20]{diaz2020interactive}
\textsc{D{\'\i}az J., Marton F., Gobbetti E.}:
\newblock Interactive spatio-temporal exploration of massive time-varying
  rectilinear scalar volumes based on a variable bit-rate sparse representation
  over learned dictionaries.
\newblock \emph{Computers \& Graphics 88} (2020), 45--56.

\bibitem[DNJ20]{davies2020effectiveness}
\textsc{Davies T., Nowrouzezahrai D., Jacobson A.}:
\newblock On the effectiveness of weight-encoded neural implicit 3d shapes.
\newblock \emph{arXiv preprint} (2020).
\newblock \href {http://arxiv.org/abs/2009.09808} {\path{arXiv:2009.09808}}.

\bibitem[EUT19]{eckert2019Scalarflow}
\textsc{Eckert M.-L., Um K., Thuerey N.}:
\newblock {ScalarFlow}: A large-scale volumetric data set of real-world scalar
  transport flows for computer animation and machine learning.
\newblock \emph{ACM Trans. Graph. 38}, 6 (Nov. 2019).

\bibitem[Fen03]{fenney2003texture}
\textsc{Fenney S.}:
\newblock Texture compression using low-frequency signal modulation.
\newblock In \emph{Proceedings of the ACM SIGGRAPH/EUROGRAPHICS conference on
  Graphics hardware} (2003), pp.~84--91.

\bibitem[FM07]{fout2007transform}
\textsc{Fout N., Ma K.-L.}:
\newblock Transform coding for hardware-accelerated volume rendering.
\newblock \emph{IEEE Transactions on Visualization and Computer Graphics 13}, 6
  (2007), 1600--1607.

\bibitem[Gav20]{gavrilescu2020supervised}
\textsc{Gavrilescu M.}:
\newblock A supervised generative model for efficient rendering of medical
  volume data.
\newblock In \emph{2020 International Conference on e-Health and Bioengineering
  (EHB)} (2020), IEEE, pp.~1--4.

\bibitem[GG16]{guthe2016variable}
\textsc{Guthe S., Goesele M.}:
\newblock Variable length coding for gpu-based direct volume rendering.
\newblock In \emph{Proceedings of the Conference on Vision, Modeling and
  Visualization} (2016), pp.~77--84.

\bibitem[GIGM12]{gobbetti2012covra}
\textsc{Gobbetti E., Iglesias~Guiti{\'a}n J.~A., Marton F.}:
\newblock {COVRA}: A compression-domain output-sensitive volume rendering
  architecture based on a sparse representation of voxel blocks.
\newblock In \emph{Computer Graphics Forum} (2012), vol.~31, Wiley Online
  Library, pp.~1315--1324.

\bibitem[GKJ{\etalchar{*}}21]{garbin2021fastnerf}
\textsc{Garbin S.~J., Kowalski M., Johnson M., Shotton J., Valentin J.}:
\newblock {FastNeRF}: High-fidelity neural rendering at 200fps.
\newblock \emph{arXiv preprint} (2021).
\newblock \href {http://arxiv.org/abs/2103.10380} {\path{arXiv:2103.10380}}.

\bibitem[GYH{\etalchar{*}}20]{guo2020ssr}
\textsc{Guo L., Ye S., Han J., Zheng H., Gao H., Chen D.~Z., Wang J.-X., Wang
  C.}:
\newblock {SSR-VFD}: Spatial super-resolution for vector field data analysis
  and visualization.
\newblock In \emph{2020 IEEE Pacific Visualization Symposium (PacificVis)}
  (2020), IEEE Computer Society, pp.~71--80.

\bibitem[HSB{\etalchar{*}}21]{Hoang2021EfficientAF}
\textsc{Hoang D. T.~A., Summa B., Bhatia H., Lindstrom P., Klacansky P., Usher
  W., Bremer P.-T., Pascucci V.}:
\newblock Efficient and flexible hierarchical data layouts for a unified
  encoding of scalar field precision and resolution.
\newblock \emph{IEEE Transactions on Visualization and Computer Graphics 27}
  (2021), 603--613.

\bibitem[HW19]{han2019tsr}
\textsc{Han J., Wang C.}:
\newblock {TSR-TVD}: Temporal super-resolution for time-varying data analysis
  and visualization.
\newblock \emph{IEEE transactions on visualization and computer graphics 26}, 1
  (2019), 205--215.

\bibitem[HW20]{han2020ssr}
\textsc{Han J., Wang C.}:
\newblock {SSR-TVD}: Spatial super-resolution for time-varying data analysis
  and visualization.
\newblock \emph{IEEE Transactions on Visualization and Computer Graphics}
  (2020).

\bibitem[HZCW21]{han2021stnet}
\textsc{Han J., Zheng H., Chen D.~Z., Wang C.}:
\newblock {STNet}: An end-to-end generative framework for synthesizing
  spatiotemporal super-resolution volumes.
\newblock \emph{IEEE Transactions on Visualization and Computer Graphics}
  (2021), 1--1.

\bibitem[INH99]{iourcha1999system}
\textsc{Iourcha K., Nayak K., Hong Z.}:
\newblock System and method for fixed-rate block-based image compression with
  inferred pixel values, 1999.
\newblock US Patent 5,956,431.

\bibitem[KWTM03]{kindlmann2003curvature}
\textsc{Kindlmann G., Whitaker R., Tasdizen T., Moller T.}:
\newblock Curvature-based transfer functions for direct volume rendering:
  Methods and applications.
\newblock In \emph{IEEE Visualization, 2003. VIS 2003.} (2003), IEEE,
  pp.~513--520.

\bibitem[LCA08]{lee2008fast}
\textsc{Lee M.-C., Chan R.~K., Adjeroh D.~A.}:
\newblock Fast three-dimensional discrete cosine transform.
\newblock \emph{SIAM Journal on Scientific Computing 30}, 6 (2008), 3087--3107.

\bibitem[Lev88]{levoy1988display}
\textsc{Levoy M.}:
\newblock Display of surfaces from volume data.
\newblock \emph{IEEE Computer graphics and Applications 8}, 3 (1988), 29--37.

\bibitem[LHU20]{Liu2020Snake}
\textsc{Liu Z., Hartwig T., Ueda M.}:
\newblock Neural networks fail to learn periodic functions and how to fix it.
\newblock \emph{arXiv preprint abs/2006.08195} (2020).
\newblock \href {http://arxiv.org/abs/2006.08195} {\path{arXiv:2006.08195}}.

\bibitem[LJLB21]{lu2021compressive}
\textsc{Lu Y., Jiang K., Levine J.~A., Berger M.}:
\newblock Compressive neural representations of volumetric scalar fields.
\newblock \emph{Computer Graphics Forum} (2021).

\bibitem[LJM21]{lei2021learning}
\textsc{Lei J., Jia K., Ma Y.}:
\newblock Learning and meshing from deep implicit surface networks using an
  efficient implementation of analytic marching.
\newblock \emph{arXiv preprint} (2021).
\newblock \href {http://arxiv.org/abs/2106.10031} {\path{arXiv:2106.10031}}.

\bibitem[MAG19]{marton2019compression}
\textsc{Marton F., Agus M., Gobbetti E.}:
\newblock A framework for gpu-accelerated exploration of massive time-varying
  rectilinear scalar volumes.
\newblock \emph{Computer Graphics Forum 38}, 3 (2019), 53--66.
\newblock \href {https://doi.org/10.1111/cgf.13671}
  {\path{doi:10.1111/cgf.13671}}.

\bibitem[MDCL{\etalchar{*}}18]{markidis2018nvidia}
\textsc{Markidis S., Der~Chien S.~W., Laure E., Peng I.~B., Vetter J.~S.}:
\newblock Nvidia tensor core programmability, performance \& precision.
\newblock In \emph{2018 IEEE International Parallel and Distributed Processing
  Symposium Workshops (IPDPSW)} (2018), IEEE, pp.~522--531.

\bibitem[MLL{\etalchar{*}}21]{martel2021acorn}
\textsc{Martel J. N.~P., Lindell D.~B., Lin C.~Z., Chan E.~R., Monteiro M.,
  Wetzstein G.}:
\newblock Acorn: adaptive coordinate networks for neural scene representation.
\newblock \emph{ACM Transactions on Graphics 40}, 4 (Aug 2021), 1--13.

\bibitem[MON{\etalchar{*}}19]{mescheder2019occupancy}
\textsc{Mescheder L., Oechsle M., Niemeyer M., Nowozin S., Geiger A.}:
\newblock Occupancy networks: Learning 3d reconstruction in function space.
\newblock In \emph{Proceedings of the IEEE/CVF Conference on Computer Vision
  and Pattern Recognition} (2019), pp.~4460--4470.

\bibitem[MRNK21]{mueller2021realtime}
\textsc{M\"{u}ller T., Rousselle F., Nov\'{a}k J., Keller A.}:
\newblock Real-time neural radiance caching for path tracing.
\newblock \emph{ACM Trans. Graph. 40}, 4 (Aug. 2021), 36:1--36:16.
\newblock \href {https://doi.org/10.1145/3450626.3459812}
  {\path{doi:10.1145/3450626.3459812}}.

\bibitem[MST{\etalchar{*}}20]{mildenhall2020nerf}
\textsc{Mildenhall B., Srinivasan P.~P., Tancik M., Barron J.~T., Ramamoorthi
  R., Ng R.}:
\newblock {NeRF}: Representing scenes as neural radiance fields for view
  synthesis.
\newblock In \emph{European conference on computer vision} (2020), Springer,
  pp.~405--421.

\bibitem[NLP{\etalchar{*}}12]{nystad2012adaptive}
\textsc{Nystad J., Lassen A., Pomianowski A., Ellis S., Olson T.}:
\newblock Adaptive scalable texture compression.
\newblock In \emph{Proceedings of the Fourth ACM SIGGRAPH/Eurographics
  Conference on High-Performance Graphics} (2012), pp.~105--114.

\bibitem[PCPMMN21]{pumarola2021d}
\textsc{Pumarola A., Corona E., Pons-Moll G., Moreno-Noguer F.}:
\newblock {D-NeRF}: Neural radiance fields for dynamic scenes.
\newblock In \emph{Proceedings of the IEEE/CVF Conference on Computer Vision
  and Pattern Recognition} (2021), pp.~10318--10327.

\bibitem[PFS{\etalchar{*}}19]{park2019deepsdf}
\textsc{Park J.~J., Florence P., Straub J., Newcombe R., Lovegrove S.}:
\newblock {DeepSDF}: Learning continuous signed distance functions for shape
  representation.
\newblock In \emph{Proceedings of the IEEE/CVF Conference on Computer Vision
  and Pattern Recognition} (2019), pp.~165--174.

\bibitem[PGM{\etalchar{*}}19]{PyTorch2019}
\textsc{Paszke A., Gross S., Massa F., Lerer A., Bradbury J., Chanan G.,
  Killeen T., Lin Z., Gimelshein N., Antiga L., Desmaison A., Kopf A., Yang E.,
  DeVito Z., Raison M., Tejani A., Chilamkurthy S., Steiner B., Fang L., Bai
  J., Chintala S.}:
\newblock {PyTorch}: An imperative style, high-performance deep learning
  library.
\newblock In \emph{Advances in Neural Information Processing Systems 32},
  Wallach H., Larochelle H., Beygelzimer A., d\textquotesingle Alch\'{e}-Buc
  F., Fox E., Garnett R., (Eds.). Curran Associates, Inc., 2019,
  pp.~8024--8035.

\bibitem[RGG{\etalchar{*}}13]{rodriguez2013survey}
\textsc{Rodriguez M.~B., Gobbetti E., Guiti{\'a}n J. A.~I., Makhinya M., Marton
  F., Pajarola R., Suter S.~K.}:
\newblock A survey of compressed gpu-based direct volume rendering.
\newblock In \emph{Eurographics (State of the Art Reports)} (2013),
  pp.~117--136.

\bibitem[RTW13]{reichl2013visualization}
\textsc{Reichl F., Treib M., Westermann R.}:
\newblock Visualization of big sph simulations via compressed octree grids.
\newblock In \emph{2013 IEEE International Conference on Big Data} (2013),
  IEEE, pp.~71--78.

\bibitem[SB21]{sahoo2021vectorsuperres}
\textsc{Sahoo S., Berger M.}:
\newblock Integration-aware vector field super resolution.
\newblock In \emph{EuroVis 2021 - Short Papers} (2021), Agus M., Garth C.,
  Kerren A., (Eds.), The Eurographics Association.

\bibitem[SDZ{\etalchar{*}}21]{srinivasan2021nerv}
\textsc{Srinivasan P.~P., Deng B., Zhang X., Tancik M., Mildenhall B., Barron
  J.~T.}:
\newblock {NeRV}: Neural reflectance and visibility fields for relighting and
  view synthesis.
\newblock In \emph{Proceedings of the IEEE/CVF Conference on Computer Vision
  and Pattern Recognition} (2021), pp.~7495--7504.

\bibitem[SW03]{Schneider2003compressiondomain}
\textsc{Schneider J., Westermann R.}:
\newblock Compression domain volume rendering.
\newblock In \emph{IEEE Visualization, 2003. VIS 2003.} (2003), pp.~293--300.

\bibitem[SZW19]{sitzmann2019scene}
\textsc{Sitzmann V., Zollh{\"o}fer M., Wetzstein G.}:
\newblock Scene representation networks: Continuous 3d-structure-aware neural
  scene representations.
\newblock \emph{arXiv preprint} (2019).
\newblock \href {http://arxiv.org/abs/1906.01618} {\path{arXiv:1906.01618}}.

\bibitem[TBR{\etalchar{*}}12]{treib2012turbulence}
\textsc{Treib M., B{\"u}rger K., Reichl F., Meneveau C., Szalay A., Westermann
  R.}:
\newblock Turbulence visualization at the terascale on desktop pcs.
\newblock \emph{IEEE Transactions on Visualization and Computer Graphics 18},
  12 (2012), 2169--2177.

\bibitem[TFT{\etalchar{*}}20]{tewari2020state}
\textsc{Tewari A., Fried O., Thies J., Sitzmann V., Lombardi S., Sunkavalli K.,
  Martin-Brualla R., Simon T., Saragih J., Nie{\ss}ner M., et~al.}:
\newblock State of the art on neural rendering.
\newblock In \emph{Computer Graphics Forum} (2020), vol.~39, Wiley Online
  Library, pp.~701--727.

\bibitem[TLY{\etalchar{*}}21]{takikawa2021neural}
\textsc{Takikawa T., Litalien J., Yin K., Kreis K., Loop C., Nowrouzezahrai D.,
  Jacobson A., McGuire M., Fidler S.}:
\newblock Neural geometric level of detail: Real-time rendering with implicit
  3d shapes.
\newblock In \emph{Proceedings of the IEEE/CVF Conference on Computer Vision
  and Pattern Recognition} (2021), pp.~11358--11367.

\bibitem[TRAW12]{treib2012interactive}
\textsc{Treib M., Reichl F., Auer S., Westermann R.}:
\newblock Interactive editing of {GigaSample} terrain fields.
\newblock In \emph{Computer graphics forum} (2012), vol.~31, Wiley Online
  Library, pp.~383--392.

\bibitem[TSM{\etalchar{*}}20]{tancik2020fourier}
\textsc{Tancik M., Srinivasan P.~P., Mildenhall B., Fridovich-Keil S., Raghavan
  N., Singhal U., Ramamoorthi R., Barron J.~T., Ng R.}:
\newblock Fourier features let networks learn high frequency functions in low
  dimensional domains.
\newblock \emph{arXiv preprint} (2020).
\newblock \href {http://arxiv.org/abs/2006.10739} {\path{arXiv:2006.10739}}.

\bibitem[WBSS04]{wang2004image}
\textsc{Wang Z., Bovik A.~C., Sheikh H.~R., Simoncelli E.~P.}:
\newblock Image quality assessment: from error visibility to structural
  similarity.
\newblock \emph{IEEE transactions on image processing 13}, 4 (2004), 600--612.

\bibitem[WCTW19]{weiss2019isosuperres}
\textsc{{Weiss} S., {Chu} M., {Thuerey} N., {Westermann} R.}:
\newblock Volumetric isosurface rendering with deep learning-based
  super-resolution.
\newblock \emph{IEEE Transactions on Visualization and Computer Graphics}
  (2019), 1--1.

\bibitem[Wes95]{westermann1995compression}
\textsc{Westermann R.}:
\newblock Compression domain rendering of time-resolved volume data.
\newblock In \emph{Proceedings Visualization '95} (1995), pp.~168--175.

\bibitem[WITW20]{weiss2020adaptivesampling}
\textsc{Weiss S., I\c{s}\i{}k M., Thies J., Westermann R.}:
\newblock Learning adaptive sampling and reconstruction for volume
  visualization.
\newblock \emph{IEEE Transactions on Visualization and Computer Graphics}
  (2020), 1--1.

\bibitem[WSG{\etalchar{*}}21]{wurster2021deep}
\textsc{Wurster S.~W., Shen H.-W., Guo H., Peterka T., Raj M., Xu J.}:
\newblock Deep hierarchical super-resolution for scientific data reduction and
  visualization, 2021.
\newblock \href {http://arxiv.org/abs/2107.00462} {\path{arXiv:2107.00462}}.

\bibitem[YFKT{\etalchar{*}}21]{yu2021plenoxels}
\textsc{Yu A., Fridovich-Keil S., Tancik M., Chen Q., Recht B., Kanazawa A.}:
\newblock Plenoxels: Radiance fields without neural networks.
\newblock \emph{arXiv preprint arXiv:2112.05131} (2021).

\bibitem[YL95]{yeo1995volume}
\textsc{Yeo B.-L., Liu B.}:
\newblock Volume rendering of dct-based compressed 3d scalar data.
\newblock \emph{IEEE Transactions on Visualization and Computer Graphics 1}, 1
  (1995), 29--43.

\bibitem[YLT{\etalchar{*}}21]{yu2021plenoctrees}
\textsc{Yu A., Li R., Tancik M., Li H., Ng R., Kanazawa A.}:
\newblock Plenoctrees for real-time rendering of neural radiance fields.
\newblock In \emph{Proceedings of the IEEE/CVF International Conference on
  Computer Vision} (2021), pp.~5752--5761.

\bibitem[ZDL{\etalchar{*}}20]{kai2020sz}
\textsc{Zhao K., Di S., Liang X., Li S., Tao D., Chen Z., Cappello F.}:
\newblock Significantly improving lossy compression for hpc datasets with
  second-order prediction and parameter optimization.
\newblock In \emph{Proceedings of the 29th International Symposium on
  High-Performance Parallel and Distributed Computing} (New York, NY, USA,
  2020), HPDC '20, Association for Computing Machinery, pp.~89--100.

\bibitem[ZHW{\etalchar{*}}17]{zhou2017volume}
\textsc{Zhou Z., Hou Y., Wang Q., Chen G., Lu J., Tao Y., Lin H.}:
\newblock Volume upscaling with convolutional neural networks.
\newblock In \emph{Proceedings of the Computer Graphics International
  Conference} (2017), pp.~1--6.

\bibitem[ZIE{\etalchar{*}}18]{zhang2018perceptual}
\textsc{Zhang R., Isola P., Efros A.~A., Shechtman E., Wang O.}:
\newblock The unreasonable effectiveness of deep features as a perceptual
  metric.
\newblock In \emph{CVPR} (2018).

\end{thebibliography}

\clearpage


\section*{Supplementary material}

# Fast Neural Representations for Direct Volume Rendering – Supplementary

Sebastian Weiss, Philipp Hermüller, and Rüdiger Westermann

**Abstract**

In this supplementary document, we provide additional experiments on alternate architectures and further results on novel datasets. First, a comparison of various activation functions is shown in Sec. 1. Training the network from images instead of from points in 3D is analyzed in Sec. 2. Further details on how to extend the network to provide gradients as well as densities are provided in Sec. 3. Two further datasets for comparisons against baseline methods are shown in Sec. 4

## 1 Activation Functions

In Section 5 of the main paper, we proposed to use an alternative version of the Snake activation function [LHU20], defined as

$$SnakeAlt(x) = \frac{x + 1 - cos(2x)}{2} = 0.5x + sin^2(x). \tag{1}$$

This function is motivated by providing a more monotonous slope than the original Snake. A comparison of various activation functions, traditional ReLU, Sine [SMB*20], Snake [LHU20], SnakeAlt, is presented in Tab. 1. As one can see, Snake and SnakeAlt are almost on par, with SnakeAlt leading for two out of three datasets.

Table 1: Evaluation of different activation functions on rendered images.

| Activation | Plume | | Ejecta | | RM | |
|---|---|---|---|---|---|---|
| | SSIM ↑ | LPIPS ↓ | SSIM ↑ | LPIPS ↓ | SSIM ↑ | LPIPS ↓ |
| ReLU | 0.9792 | 0.0261 | 0.8912 | 0.1024 | 0.8244 | 0.1434 |
| Sine | 0.9760 | 0.0208 | 0.8784 | 0.1312 | 0.8166 | 0.1490 |
| Snake | 0.9867 | 0.0104 | **0.9102** | **0.0872** | 0.8438 | 0.1317 |
| SnakeAlt | **0.9893** | **0.0100** | 0.9094 | 0.0920 | **0.8458** | **0.1275** |

# 2 Image-based Training

In most cases, the encoding quality of a compact model representation is measured by how well the decoded data represents the initial data. In the following, we intend to consider the encoding quality in a different way, by considering an encoding optimal if it takes into account the perceptual investigation of the encoded information when performing visual data analysis tasks. In particular, if it is critical for the user to confuse two structures because they belong to different important features, then codes should be designed so that these structures are easily discriminated, even if this means a loss of precision for other structures. As a consequence, the encoding of data that are task-irrelevant should be allocated minimal resources. This principle is fundamental to data visualization, since it asks for the reconstruction of data from a perceptual point of view. The question of how to generate such task-dependent, i.e. perceptual-adaptive, codes that can be intertwined with a renderer in the envisioned way has not been addressed in visualization. In the following, we make a first step towards this goal via the use of SRNs.

## 2.1 Training Methodology

To achieve the aforementioned goal, we train fV-SRN in screen-space instead of in world-space using position-density mappings. However, when training in screen-space the network cannot reconstruct a density field from the provided color images. Weiss and Westermann [WW22] have shown that the non-convexity of the applied TFs make the reconstruction of densities ill-posed. Therefore, the networks used for screen-space training directly predict the color and absorption. The rgb-color is parameterized to $[0, 1]$ by a sigmoid function and the absorption to $\mathbb{R}^+$ by a softplus function to restrict the values to the valid range for blending.

In the training process, pairs $(C, I)$ consisting of the camera intrinsics $C$ and the reference image $I$ are used. Images of the volume from multiple views are then rendered via raytracing using the SRN, and the network learns to adapt its internal mapping function so that the reference images are best preserved. In the original version by Sitzmann *et al.* [SZW19] and Mildenhall *et al.* [MST*20], the training method is extremely memory intensive, because the intermediate values along the rays of sight have to be stored for backpropagation. This severely limits the number of rays or steps that can be processed. For example, Mildenhall *et al.* report that only 128 steps along the ray with an image resolution of $64^2$ can be used for training. To support arbitrarily many steps along the ray during training, we use the approach proposed by Weiss and Westermann [WW22] for differentiable volume rendering. This approach exploits an analytic inversion of the blending function to ensure a constant memory footprint, regardless of the number of steps per ray that need to be backpropagated.

As described in Sec. 5.3 of the main work, Fourier features are required by the SRN to enable the encoding and accurate reconstruction of high-frequency details. When training fV-SRN in screen-space, besides position the view direction is now available as additional input. This allows for different variations of the input encoding. Let $\boldsymbol{p} = (p_x, p_y, p_z)$ be the position of the

Table 2: Quality comparison between training in world-space and in screen-space with different input encodings and stepsizes for the TF from Fig. 1.

| Input | Stepsize $\sim$ 1 voxel | | Stepsize $\sim$ 4 voxel | |
|---|---|---|---|---|
| | SSIM $\uparrow$ | LPIPS $\downarrow$ | SSIM $\uparrow$ | LPIPS $\downarrow$ |
| pos | 0.9430 | 0.0508 | 0.8589 | 0.2106 |
| dirP | 0.9482 | 0.0542 | 0.9038 | 0.1693 |
| dirF | **0.9548** | 0.0565 | 0.8926 | 0.1887 |
| world | 0.9515 | **0.0288** | **0.9116** | **0.0777** |

current sample as seen by direction $\boldsymbol{d} = (d_x, d_y, d_z)$, let $F$ be the Fourier matrix. Then the input encodings—which we further analyse in the following ablation study, see below—are as follows:

$$\boldsymbol{v}_{\text{pos}} = \boldsymbol{p} \oplus \sin(F\boldsymbol{p}) \oplus \cos(F\boldsymbol{p}) \tag{2a}$$

$$\boldsymbol{v}_{\text{dirP}} = \boldsymbol{p} \oplus \boldsymbol{d} \oplus \sin(F\boldsymbol{p}) \oplus \cos(F\boldsymbol{p}) \tag{2b}$$

$$\boldsymbol{v}_{\text{dirF}} = \boldsymbol{p} \oplus \boldsymbol{d} \oplus \sin(F(\boldsymbol{p} \oplus \boldsymbol{d})) \oplus \cos(F(\boldsymbol{p} \oplus \boldsymbol{d})) \tag{2c}$$

## 2.2 Ablation Study and Results

So far, all networks we have analyzed have been trained in world-space, i.e., domain positions and corresponding density or color samples have been used in the training process. Now, the encodings introduced in Sec. 2.1 are used during training in screen-space.

The first observation is that the times for screen-space training exceed the times for world-space training to some extent. On our target architecture, training the world-space network for 500 epochs – to ensure convergence – and 16 million points takes roughly 30 minutes. Screen-space training with a stepsize of 0.02, 96 images of resolution $256^2$, i.e. 6 million pixels, on the other hand, takes about 2½ hours. In the following, we further analyze the resulting reconstruction quality of color prediction.

As can be seen in Tab. 2, for a small stepsize of $\sim$ 1 voxel all networks show almost equal quality, with a slight advantage of world-space training. Only the results for a TF with sharp peaks are shown. For TFs with smoother peaks, the results behave similarly. These results are surprising since we were expecting an advantage of screen-space training because the networks' capacities should have been dedicated to those structures being seen in the images.

Even more interesting are the qualitative results shown in Fig. 1. For the TF with sharp peaks and a stepsize of $\sim$ 4 voxels, ringing artifacts occur due to undersampling when the network is trained in world-space. The networks trained in screen-space do not show such artifacts, since they are trained from artifact-free reference images. However, these networks have difficulties in accurately recovering sharp color transitions in the images and tend to produce blurry results. On the other hand, even at a rather coarse stepsize the overall structure of the rendered field is reconstructed since the network seems to use its capacities on those structures that are seen in

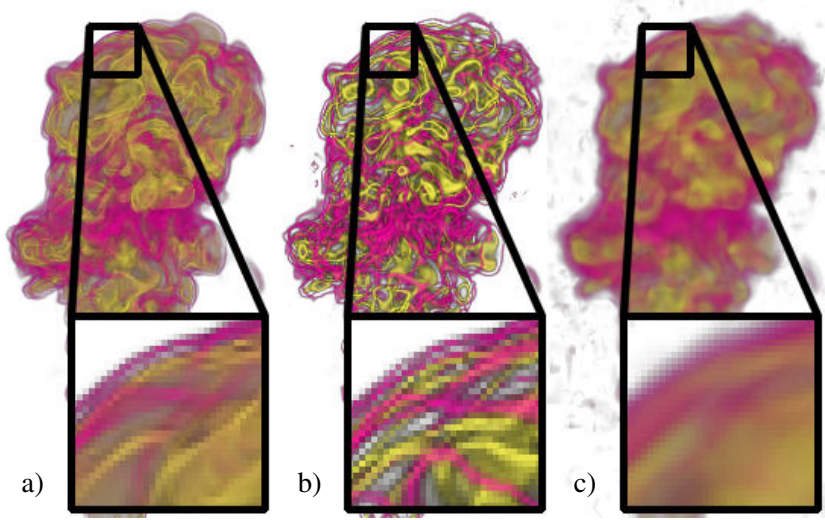

Figure 1: Screen- vs. world-space training, using a TF with sharp peaks. a) Reference rendering with stepsize of 0.1 voxels. b) Rendering with a stepsize of 4 voxels and reconstruction using fV-SRN trained in world-space. c) Rendering with a stepsize of 4 voxels and reconstruction using fV-SRN trained in screen-space.

the reference images. Due to these observations, we believe that it is worth analyzing screen-space training further and shed light on how to extend the network capacities so that accurate reconstruction becomes possible.

## 3 Gradient-predicting SRNs

In this section, we extend upon the discussion of SRNs with gradient prediction. First, in Sec. 3.1 the possible options for evaluating gradients with existing SRNs or extending those are discussed. Extending SRNs to directly predict gradients requires retraining and involves further hyperparameters, which are discussed in Sec. 3.2. Comparisons on implicit functions and on novel datasets are conducted in Sec. 3.3.

### 3.1 Methods for Computing Gradients

Let $f_\Theta : \mathbb{R}^3 \to \mathbb{R}$ be the fV-SRN function that maps from position $\boldsymbol{p} \in \mathbb{R}^3$ to density $d \in \mathbb{R}$. We now discuss how to compute density gradients $\nabla f_\Theta$ using finite differences (Sec. 3.1.1) or automatic differentiation (Sec. 3.1.2). Those approaches act directly on the density-predicting SRN without the need to change the architecture but introduce additional memory and/or computational overhead. Alternatively, how to train the networks to directly predict the density and density gradient is presented in Sec. 3.1.3. As shown below, this approach promises gradient prediction without additional memory and computation cost.

When comparing the various approaches, we analyze the cost of the various methods using two values: $C \in \mathbb{N}$ being the number of additional network evaluations required to compute the

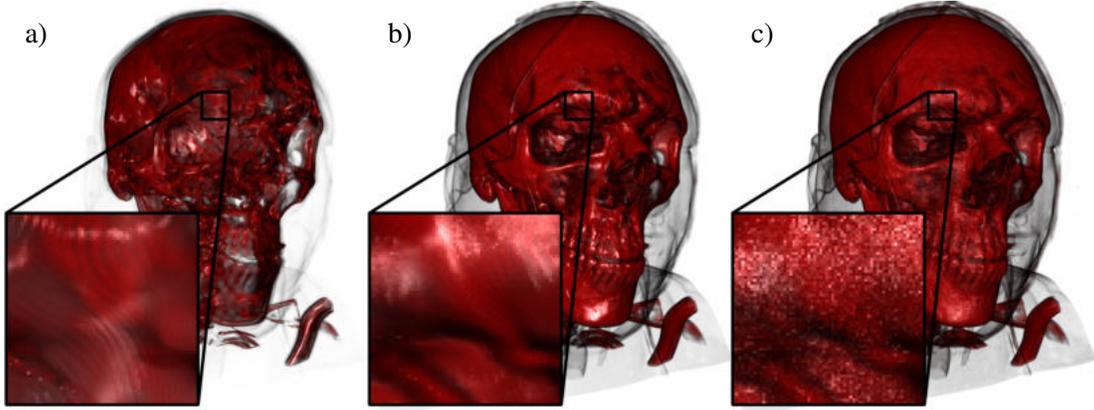

Figure 2: Renderings with a gradient-depending TF with a different stepsize $h$ for the finite differences approximation: a) $h = 4R$, b) $h = R$, c) $h = R/16$, where $R$ is the voxel size.

gradient, and $M \in \mathbb{N}$ being the number of hidden state vectors $\boldsymbol{v}_i$ that have to be stored. A higher value of $M$ indicates possible lower performance, as fewer threads can be executed in parallel so as to not exceed the available shared memory and registers on the GPU.

### 3.1.1 Finite Differences

Using central differences, the gradient is approximated as [Lev88]

$$\nabla f_\Theta(x, y, z) = \frac{1}{2h} \begin{pmatrix} f_\Theta(x+h, y, z) - f_\Theta(x-h, y, z) \\ f_\Theta(x, y+h, z) - f_\Theta(x, y-h, z) \\ f_\Theta(x, y, z+h) - f_\Theta(x, y, z-h) \end{pmatrix}, \tag{3}$$

where $h$ is the stepsize of the finite differences. Choosing $h$ too large (Fig. 2a) may lead to incorrect gradients as small features are jumped over. Choosing $h$ too small (Fig. 2c) leads to numerical instabilities due to cancellation of significant bits in the subtraction. When a regular, trilinearly interpolated hexahedral grid is approximated with the SRN, we can choose $h$ optimally to be equal to the voxel size (Fig. 2b). In total, this approach requires six additional network evaluations, hence $C = 6$, but no additional hidden states have to be stored, $M = 0$.

### 3.1.2 Automatic Differentiation

The SRN can be written as a sequence of vector-valued functions

$$\boldsymbol{y} = f(g(h(\boldsymbol{x}))). \tag{4}$$

Then the gradient of that sequence of functions at position $\boldsymbol{x}$ can be expressed using the Jacobian matrices,

$$J_{f \circ g \circ h}(\boldsymbol{x}) = J_f(g(h(\boldsymbol{x}))) J_g(h(\boldsymbol{x})) J_h(\boldsymbol{x}). \tag{5}$$

There exist multiple ways how to evaluate the resulting matrix products, all of which are, apart from numerical rounding errors, identical. Evaluating the product left-to-right leads to the Adjoint

Method, right-to-left evaluation leads to forward differentiation. For a more detailed revision of analytic differentiation for rendering, we refer to [WW22].

**Adjoint Method**    In the adjoint method [MTPS04], gradients for a variable $\boldsymbol{x}$ are traced via the so-called adjoint variables $\hat{\boldsymbol{x}}$ that store the gradient of that variable with respect to the scalar output, i.e. the density. This computation is performed after the regular network evaluation in the reversed order of operations. For an operation $\boldsymbol{y} = f(\boldsymbol{x})$, the gradients are propagated from the output to the input via $\hat{\boldsymbol{x}} = J_f(\boldsymbol{x})^T \hat{\boldsymbol{y}}$. In the concrete case of SRNs, the derivatives for a layer evaluation and activation function are given as

$$\boldsymbol{y} = W\boldsymbol{x} + b \Rightarrow \hat{\boldsymbol{x}} = W^T \hat{\boldsymbol{y}} \tag{6}$$

$$y = SnakeAlt(x) \Rightarrow \hat{x} = (0.5 - \sin(2x))\hat{y}. \tag{7}$$

Note that only the activation function – being a non-linear function – requires the input value for computing the derivative. Therefore, the hidden state before the *SnakeAlt* activation function has to be stored in the regular density evaluation for later use. With $L = 4$ layers, where the last layer only has an identity activation function, $M = 3$ hidden state vectors have to be stored. Computation-wise, Eq. 6 is equal to a regular network evaluation, hence only $C = 1$ additional network evaluations are needed.

**Forward Differentiation**    As opposed to the adjoint method, in forward differentiation [Nei10, BBBCD00], the gradients are propagated jointly together with the density estimate in the following way: Each variable $\boldsymbol{x} \in \mathbb{R}^d$ is appended by partial derivatives with respect to $x, y, z$ of the input, $\tilde{\boldsymbol{x}} = \langle \boldsymbol{x}, \boldsymbol{x}_x, \boldsymbol{x}_y, \boldsymbol{x}_z \rangle$, where $\boldsymbol{x}_x := \mathrm{d}\boldsymbol{x}/\mathrm{d}x$ and $\boldsymbol{x}_y, \boldsymbol{x}_z$ follow similarly. The input position is initialized as

$$\tilde{\boldsymbol{p}} = \langle [x, y, z]^T, [1, 0, 0]^T, [0, 1, 0]^T, [0, 0, 1]^T \rangle, \tag{8}$$

and a function $\boldsymbol{y} = f(\boldsymbol{x})$ is transformed into

$$\tilde{\boldsymbol{y}} = \tilde{f}(\tilde{\boldsymbol{x}}) = \langle f(\boldsymbol{x}), J_f(\boldsymbol{x})\boldsymbol{x}_x, J_f(\boldsymbol{x})\boldsymbol{x}_y, J_f(\boldsymbol{x})\boldsymbol{x}_z \rangle. \tag{9}$$

After the last operation, the so-called forward variable $\tilde{d}$ of the density contains the density value itself and the derivatives with respect to the $x, y, z$-axis. As opposed to the adjoint method, the derivatives are propagated jointly with the hidden values and don't have to be saved for later. However, the storage of the forward variables for the hidden states required additional $M = 3$ memory. Similarly, $C = 3$ additional matrix evaluations are needed.

Note that the performance scores $M, C$ are independent of the network architecture for the forward method. For the adjoint method, however, the memory requirements $M$ scale with the number of layers in the network. Therefore, if larger networks provide an issue memory-wise with the adjoint method, the forward method becomes an alternative. For the fV-SRN architecture with only four layers, however, the adjoint method is always advantageous.

### 3.1.3  Direct Gradient-Predicting SRNs

As shown above, both finite differences ($C = 6, M = 0$) and the adjoint method ($C = 1, M = 3$) have a computational and/or memory overhead. We show, that it is possible to train SRNs to

predict the density gradient alongside the regular density output. This introduces no additional inference cost ($C = 0, M = 0$). To that end, we modify the fV-SRN architecture to predict four scalar values as output, $d$ and $\nabla d$. As loss function between the network prediction $\cdot_{\text{pred}}$ and the ground truth reference $\cdot_{\text{gt}}$, we employ a combination of losses on the density and the gradient,

$$\mathcal{L} = (1 - \alpha)|d_{\text{pred}} - d_{\text{gt}}| + \alpha ||\nabla d_{\text{pred}} - \nabla d_{\text{gt}}||_2^2. \tag{10}$$

We train the network on triples of position, density, and gradient, where the gradient is computed analytically for implicit functions (Sec. 3.3.1) or via finite differences on a grid (Sec. 3.3.2). All other training parameters stay the same.

## 3.2 Hyperparameters and Ablation Study

Introducing gradient prediction in the SRNs adds several hyperparameters to the optimization process:

1. The weighting term $\alpha$ between the loss on the densities and the loss on the gradients, see Eq. 10, has to be specified. We use the following parametrization to explore the possible space of $\alpha$, $\alpha := 0.5 \tanh(0.5w) + 0.5$, with $w \in \mathbb{Z}$.

2. $\mathcal{L}_1$ and $\mathcal{L}_2$ losses were tested as the loss function for the gradients.

3. For the output parametrization of the gradients, we experimented with using the identity function on the output of the last layer for the gradients ("densitygrad:direct"), or raising the output to the third power ("densitygrad:cubic") to help the network produce larger gradient values.

4. For resampling the training data to generate new samples in areas with high prediction errors, the error can be computed either directly on the densities and gradients ($c =$ False) or on the shaded color after applying the transfer function ($c =$ True). During resampling, each location is chosen with a minimal probability $p$ to keep a few samples in areas that are already predicted well. This prevents unlearning of those areas.

While the quality of the networks could be evaluated quantitatively by measuring, e.g., the PSNR of the predicted densities and gradients, it is unclear, how to weigh those two error scores. For example, the PSNR score on the gradients can always be improved by putting more weight on the loss function for the gradients. This, however, usually degrades the PSNR score for the densities. We, therefore, opt to analyze the quality of the networks on rendered images using the LPIPS score [ZIE*18]. This leads to a natural weighting, as a good prediction of the density is required to provide the general outline of the object, while a good prediction of the gradient is needed for proper surface shading. We apply the LPIPS score on two images. First, on a color image, where the RGB color is provided by the TF and shaded using Phong. Second, on a normal image, where the predicted gradient is normalized, linearly remapped fron $[-1, +1]^3$ to $[0, 1]^3$, treated as color and blended based on the absorption from the TF.

The results can be found in Fig. 3. As datasets, we chose a CT scan of a human skull with resolution $256^3$ and a Richtmyer-Meshkov simulation with resolution $512^3$. To assert the

variance in the training runs, for each configuration, three runs with three different seeds were launched. The scores for those single runs are drawn as small colored circles, the mean score is used to color the whole cell in the heatmaps. Several observations can be drawn from the heatmap of the LPIPS scores. First, with respect to the gradient weight $w$, regardless of the other hyperparameters, a U-shape can be observed: The prediction worsens with a too-low or too-high weighting between the density and gradient loss. The optimal value differs between an $\mathcal{L}_1$ and $\mathcal{L}_2$ because of the different scales, and also between datasets. Second, the choice of the loss function is critical. While for the skull dataset, $\mathcal{L}_1$ and $\mathcal{L}_2$ perform almost equal, for the Richtmyer-Meshkov dataset on the normal images, the $\mathcal{L}_2$ loss is advantageous. Third, the choice of output parametrization plays only a secondary role. For the Richtmyer-Meshkov dataset, both analyzed output parametrizations are almost equal, for the skull dataset, direct prediction is slightly better than "cubic". Forth, we found that the resampling strategy has no significant impact on the quality, any choice is viable. Last, we observed that the individual training runs for "good" configurations show only minor differences. This indicates that those configurations are good in general, and not a product of chance due to the random nature of network training. Configurations that show higher, i.e. worse, LPIPS scores show a much higher variance between the individual runs.

In total, there exist many configurations that lead to similar good reconstructions. To optimal settings differ from dataset to dataset slightly. For generalization to other datasets, running the whole hyperparameter study for every novel dataset is infeasible. Therefore, we decided on a single configuration, colored in green, that leads to reasonable good results for both the skull and the Richtmyer-Meshkov dataset: A gradient weighting of $w = -6, \alpha \approx 0.0025$, $\mathcal{L}_2$ loss function on the gradients, "direct" output parametrization, and resampling with $p = 0.1$ and $c =$ False. We recommend this configuration as a starting point for novel datasets. To improve the quality, the parameter $\alpha$ can be fine-tuned. A sub-optimal configuration is highlighted in blue in Fig. 3 as an example for a failure case.

## 3.3 Evaluation

We evaluate the prediction quality and inference speed of three methods: finite differences (*FD*) for its ease of implementation, the adjoint method (*Adjoint*) for being the fastest method for automatic differentiation, and direct gradient-predicting SRNs (*Direct*), that require (re-)training with gradients in mind.

With the hyperparameter study on the Skull and Richtmyer-Meshkov dataset, see Sec. 3.2 above, we found that an $\mathcal{L}_2$ loss on the gradients leads to better results than an $\mathcal{L}_1$ loss for training the direct gradient-predicting SRNs. The optimal value for $\alpha$, however, differs from dataset to dataset. A value of $\alpha = 0.0025$ provides a good compromise and is used in Sec. 3.3.1. For the novel datasets in Sec. 3.3.2, this value is fine-tuned.

### 3.3.1 Implicit Functions with Analytical Gradients

First, we evaluate the three presented methods on implicit functions, "Blobby" [SN09] and "Marschner Lobb" (ML) [ML94]. These two examples provide analytic functions for the density

Figure 3: A hyperparameter study was conducted on two datasets, a human skull ($256^3$, top) and a Richtmyer-Meshkov simulation ($512^3$, bottom). The quality of the runs was evaluated using LPIPS on the color and normal image, a lower value means a better reconstruction and is represented by a brighter color. The four evaluated hyperparameters are: the output parametrization and gradient weight $w$ along X, and the gradient loss function and resampling strategy along Y. To validate potential variance in the training, for each configuration, three runs are started with three different seeds. Those runs are depicted by the small colored circle in each cell. Hence, each circle represents a single training run. The cells are colored by the mean score. The reference images are outlined in black, the final recommended configuration is highlighted in green, and a failure configuration is highlighted in blue.

9

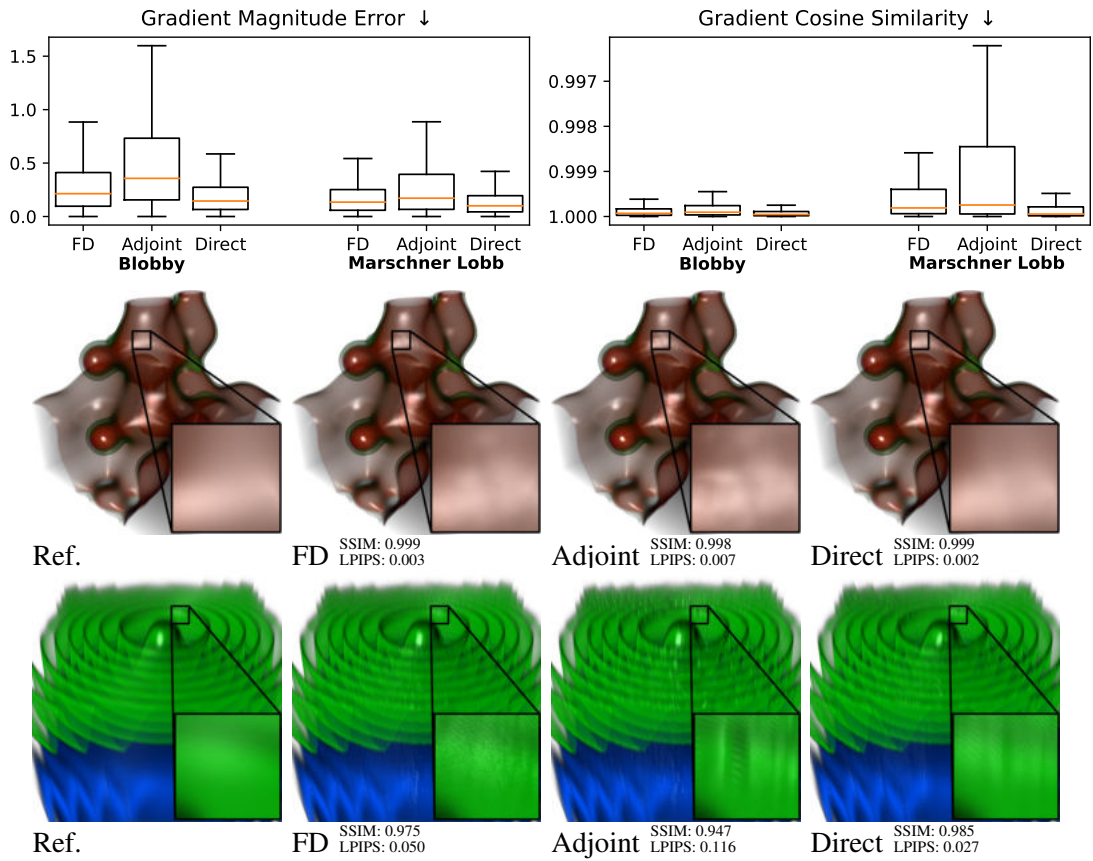

Figure 4: Evaluation of analytic datasets, "Blobby" [SN09] and "Marschner Lobb" [ML94]. The box plots show the absolute error on the predicted gradient magnitude and cosine similarity of the gradient direction. A lower value is better.

$f$ and the gradient $\nabla f$, which thus allow evaluating the presented methods without biases and errors introduced due to discretization on a grid.

For the quantitative evaluations, we focus on the errors on the gradient prediction, see Fig. 4 top. The error analysis is separated into two parts: First, the absolute error on the gradient magnitude $f'$ is plotted, as needed e.g. for transfer functions that take $f'$ as an additional axis. Second, the cosine similarity between the predicted and ground truth gradient is used, because a correct prediction of the angle of the gradient is needed for shading. As one can see from Fig. 4, the mean error for both statistics is lowest for the direct gradient-predicting SRNs (*Direct*), indicating that the gradient-predicting SRNs can successfully learn the gradients alongside the densities for these datasets. Surprisingly, pure density-SRNs with the adjoint method for gradient computation (*Adjoint*) perform the worst. If the density-SRN would learn a perfect reconstruction of $f$, the adjoint method would provide $\nabla f$ perfectly as well. In our case, we attribute the high error of *Adjoint* to numerical inaccuracies because the input values are passed through twice as many computations as, e.g., *Direct*, with, therefore, twice as many chances for rounding errors.

10

Table 3: Performance comparison of the three tested methods for computing the density and gradients for $2^{24}$ samples.

| Dataset | Forward | Fwd. w/ saving | Direct | Finite Diff. | Adjoint |
|---------|---------|----------------|--------|--------------|---------|
| Blobby | 0.056 | 0.070 (×1.25) | 0.055 (×0.97) | 0.514 (×9.16) | 0.227 (×4.04) |
| ML | 0.046 | 0.090 (×1.94) | 0.057 (×1.24) | 0.523 (×11.29) | 0.188 (×4.07) |
| RM | 0.042 | 0.070 (×1.65) | 0.059 (×1.39) | 0.510 (×12.06) | 0.237 (×5.62) |
| Jet | 0.041 | 0.073 (×1.77) | 0.056 (×1.35) | 0.514 (×12.50) | 0.200 (×4.88) |
| Skull | 0.043 | 0.072 (×1.67) | 0.059 (×1.38) | 0.509 (×11.79) | 0.189 (×4.37) |
| Thorax | 0.043 | 0.074 (×1.73) | 0.071 (×1.66) | 0.525 (×12.27) | 0.239 (×5.60) |

In the qualitative evaluations in Fig. 4, one can see that all three methods are able to reconstruct the datasets accurately with only minor differences. *Direct* leads to the smoothest results and the best scores for SSIM [WBSS04] and LPIPS [ZIE*18], except for one case of the Marschner Lobb, where *FD* achieves a slightly better SSIM score.

### 3.3.2 Real-world Datasets

Next, we evaluate the proposed methods on real-world datasets. Two datasets, a Richtmyer Meshkov (RM) simulation of resolution $512^3$ and a CT scan of a human skull of resolution $256^3$ were used for the hyperparameter study, see Sec. 3.2. Here, we show how those hyperparameters are also applicable to training on new datasets, a CT scan of a human thorax of resolution $512 \times 512 \times 284$ and a jetstream simulation [LJLB21] of resolution $512 \times 336 \times 768$, see Fig. 5. Fine-tuning the gradient weighting $\alpha$ leads to slightly better reconstruction over the default value presented in Sec. 3.3, as shown in Fig. 6.

Within the grids, trilinear interpolation is used to compute the densities. The reference gradients are evaluated using central differences on again trilinear interpolated densities with a step size equal to the voxel size [Lev88]. This introduces a bias towards *FD*, as, assuming that the density field is predicted perfectly, the same method is used to compute the gradients.

This bias shows in the results, see Fig. 5. Using *FD* leads to the best SSIM and LPIPS scores. Directly predicting the gradients jointly to the densities worsens the quality slightly, as the network has to predict additional outputs using the same latent grid and network architecture. However, the rendering is around 10× faster, as evaluated in more detail in Sec. 3.3.3. The adjoint method is situated in between.

### 3.3.3 Performance

Lastly, we evaluate the performance of the presented methods when evaluating the density and gradients for $2^{24}$ samples, see Tab. 3. Timings for rendered images of resolution $1024^2$ with a stepsize of $1/512$ through the unit cube can be found in Fig. 5. All methods were implemented using CUDA TensorCores. As a baseline, we use density prediction using SRNs without gradients (*Forward*). Using the proposed direct gradient-predicting SRNs (*Direct*) leads to an overhead of up to 1.66× the rendering time of the baseline due to the larger size of the last layer in the

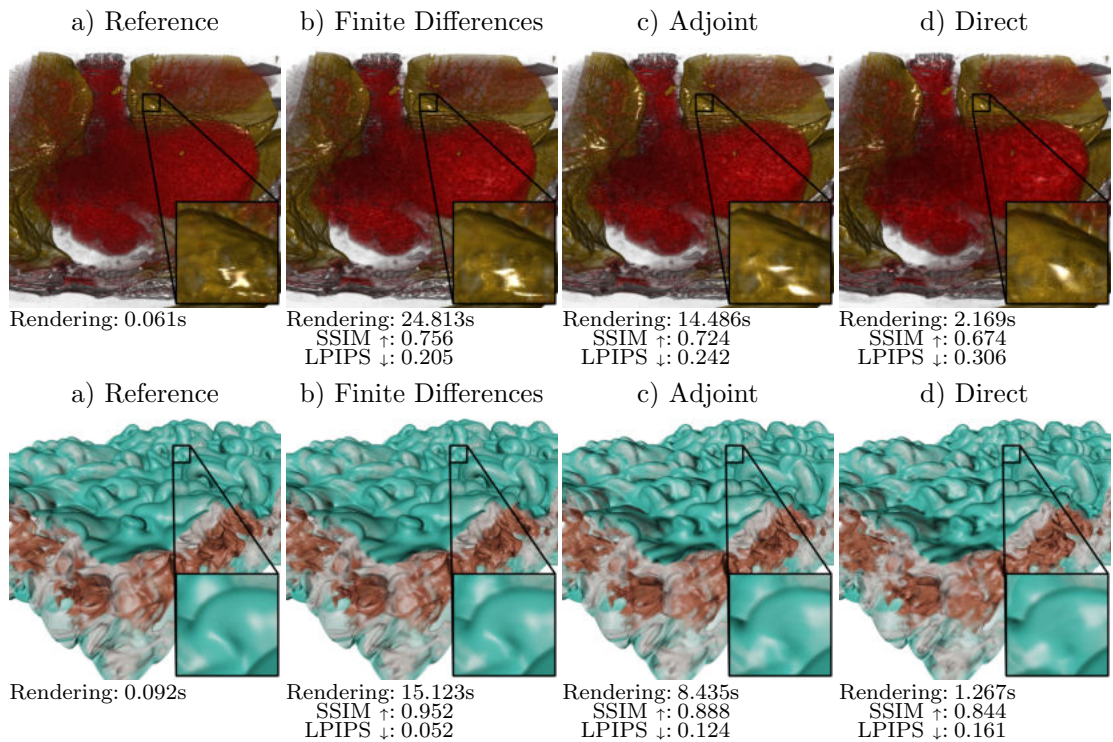

Figure 5: We explore different methods on how density gradients can be computed using volumetric SRNs, shown here on two exemplary datasets: left, a CT scan of a human thorax of resolution $512 \times 512 \times 284$, right, a jetstream simulation [LJLB21] of resolution $512 \times 336 \times 768$. SRNs allow direct rendering from a compressed representation, here with a compression ratio of $1:128$. (a) Reference rendering of the original volume using Phong shading. (b) density-predicting SRN using finite differences for the gradients, or (c) using the adjoint method. (d) extending the network to predict the gradients jointly with the density. All images were rendered to an image of resolution $1024^2$ with a fine stepsize of 1000 steps per ray on average.

network. For the adjoint method, the intermediate results of the network have to be stored in shared memory. This already decreases the performance up to $1.77\times$ (*Fwd. w/ saving*). In total, the adjoint method (*Adjoint*) is up to $5.62\times$ slower than the baseline. The slowest method uses finite differences with an overhead of up to $12.50\times$. These timings roughly follow the theoretical values from Sec. 3.1, but are in general higher than expected due to secondary effects like worse cache behavior or less parallelization due to greater register usage.

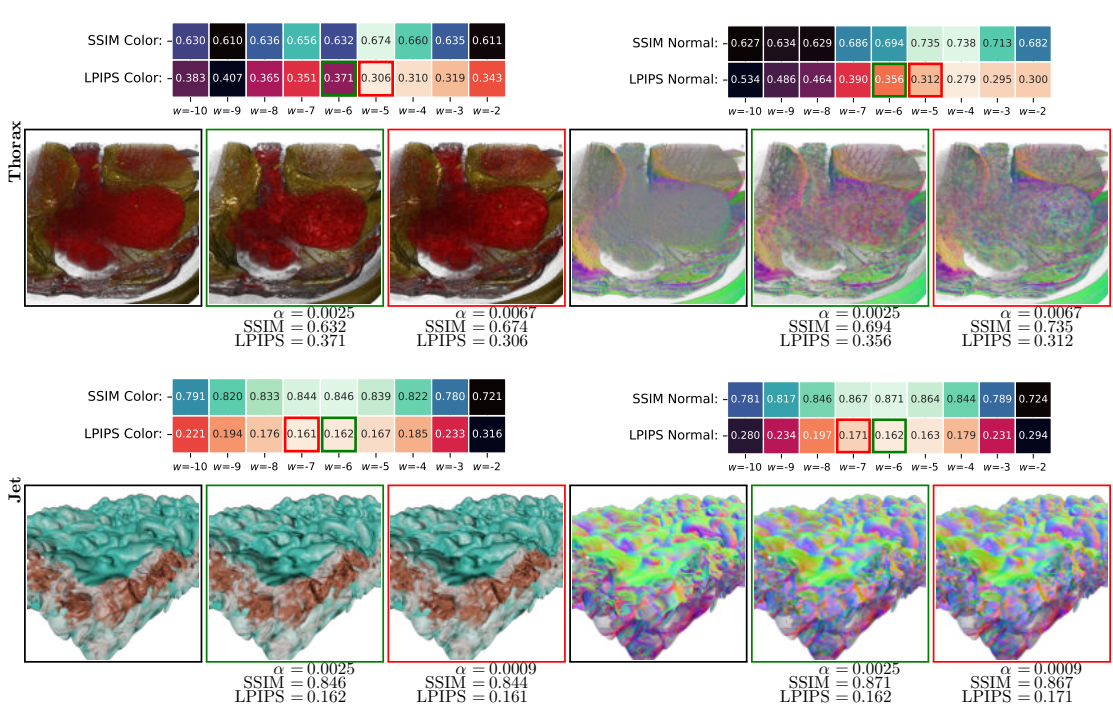

Figure 6: Influence of fine-tuning the weighting value $\alpha$ in the loss function for the datasets from Fig. 5: The default setting for the gradient weighting $\alpha$ (see Sec. 3.3) leads to the configuration highlighted in green. This parameter can be fine-tuned to produce slighly better results highlighted in red. This improved configuration is shown in Fig. 5.

## 4 Comparisons against Baseline Compression Methods – Further Results

To analyze how fV-SRN fairs against related methods, we compare ourselves against neurcomp [LJLB21], TThresh [BRLP19], and cudaCompress [TRAW12, TBR*12] with different compression ratios. The main paper includes a comparison on the Jet and Ejecta dataset, here we show two additional examples on novel datasets, a Richtmyer-Meshkov simulation of resolution $1024^3$ and the Miranda dataset of resolution $512^3$ (curtesy of Lu *et al.* [LJLB21]). The quantitative and qualitative results can be found in Fig. 7 and Fig. 8.

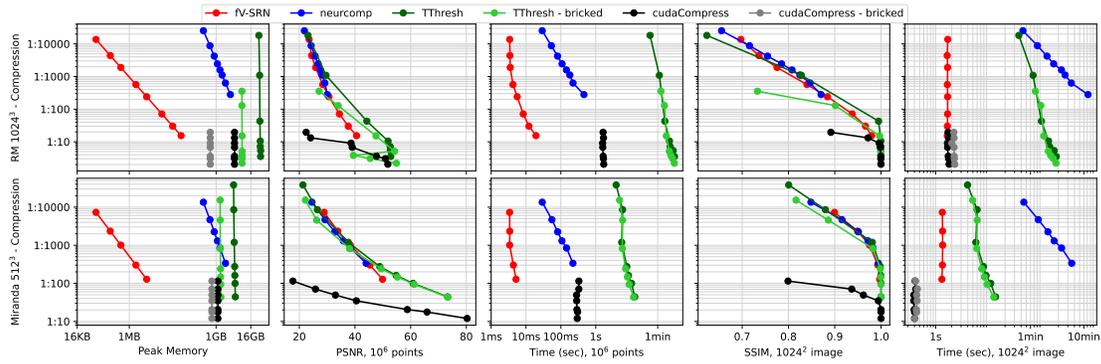

Figure 7: Evaluation of different compression rates for fV-SRN, neurcomp [LJLB21], TThresh [BRLP19], and cudaCompress [TRAW12, TBR*12], on a Richtmyer-Meshkov simulation of resolution $1024^3$ and the Miranda dataset of resolution $512^3$ (curtesy of Lu *et al.* [LJLB21]).

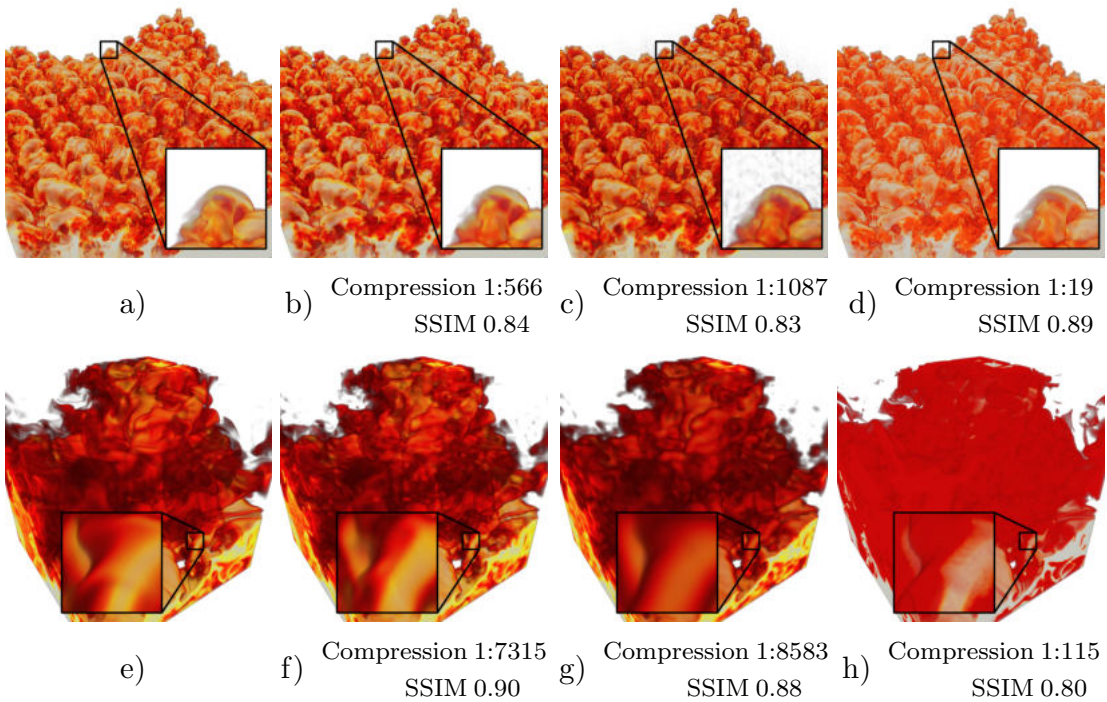

a)

b) Compression 1:566
SSIM 0.84

c) Compression 1:1087
SSIM 0.83

d) Compression 1:19
SSIM 0.89

e)

f) Compression 1:7315
SSIM 0.90

g) Compression 1:8583
SSIM 0.88

h) Compression 1:115
SSIM 0.80

Figure 8: Visual comparison of fV-SRN for different compression ratios on the Richtmyer Meshkov dataset (a-d) and the Miranda dataset (e-h). (a,e) Reference, (b,f) fV-SRN, (c,g) TThresh, (d-h) cudaCompress. For each algorithm, the example with the compression ratio closest to 1:1000 for Richtmyer Meshkov (1:10,000 for Miranda) was selected from Fig. 7.

# References

[BBBCD00] BARTHOLOMEW-BIGGS M., BROWN S., CHRISTIANSON B., DIXON L.: Automatic differentiation of algorithms. *Journal of Computational and Applied Mathematics 124*, 1 (2000), 171–190. Numerical Analysis 2000. Vol. IV: Optimization and Nonlinear Equations. `doi:10.1016/S0377-0427(00)00422-2`.

[BRLP19] BALLESTER-RIPOLL R., LINDSTROM P., PAJAROLA R.: TTHRESH: Tensor compression for multidimensional visual data. *IEEE transactions on visualization and computer graphics 26*, 9 (2019), 2891–2903.

[Lev88] LEVOY M.: Display of surfaces from volume data. *IEEE Computer graphics and Applications 8*, 3 (1988), 29–37.

[LHU20] LIU Z., HARTWIG T., UEDA M.: Neural networks fail to learn periodic functions and how to fix it. *arXiv preprint abs/2006.08195* (2020). `arXiv:2006.08195`.

[LJLB21] LU Y., JIANG K., LEVINE J. A., BERGER M.: Compressive neural representations of volumetric scalar fields. *Computer Graphics Forum* (2021).

[ML94] MARSCHNER S. R., LOBB R. J.: An evaluation of reconstruction filters for volume rendering. In *Proceedings Visualization'94* (1994), IEEE, pp. 100–107.

[MST*20] MILDENHALL B., SRINIVASAN P. P., TANCIK M., BARRON J. T., RAMAMOORTHI R., NG R.: NeRF: Representing scenes as neural radiance fields for view synthesis. In *European conference on computer vision* (2020), Springer, pp. 405–421.

[MTPS04] MCNAMARA A., TREUILLE A., POPOVIĆ Z., STAM J.: Fluid control using the adjoint method. *ACM Trans. Graph. 23*, 3 (Aug. 2004), 449–456. `doi:10.1145/1015706.1015744`.

[Nei10] NEIDINGER R. D.: Introduction to automatic differentiation and matlab object-oriented programming. *SIAM review 52*, 3 (2010), 545–563.

[SMB*20] SITZMANN V., MARTEL J. N. P., BERGMAN A. W., LINDELL D. B., WETZSTEIN G.: Implicit neural representations with periodic activation functions. *CoRR abs/2006.09661* (2020). `arXiv:2006.09661`.

[SN09] SINGH J. M., NARAYANAN P.: Real-time ray tracing of implicit surfaces on the gpu. *IEEE transactions on visualization and computer graphics 16*, 2 (2009), 261–272.

[SZW19] SITZMANN V., ZOLLHÖFER M., WETZSTEIN G.: Scene representation networks: Continuous 3d-structure-aware neural scene representations. *arXiv preprint* (2019). `arXiv:1906.01618`.



\end{document}